## DISCLAIMER

# Blueprint for DOE Quantum Supercomputing

## ENSURING U.S. LEADERSHIP IN THE QUANTUM DECADE

2026

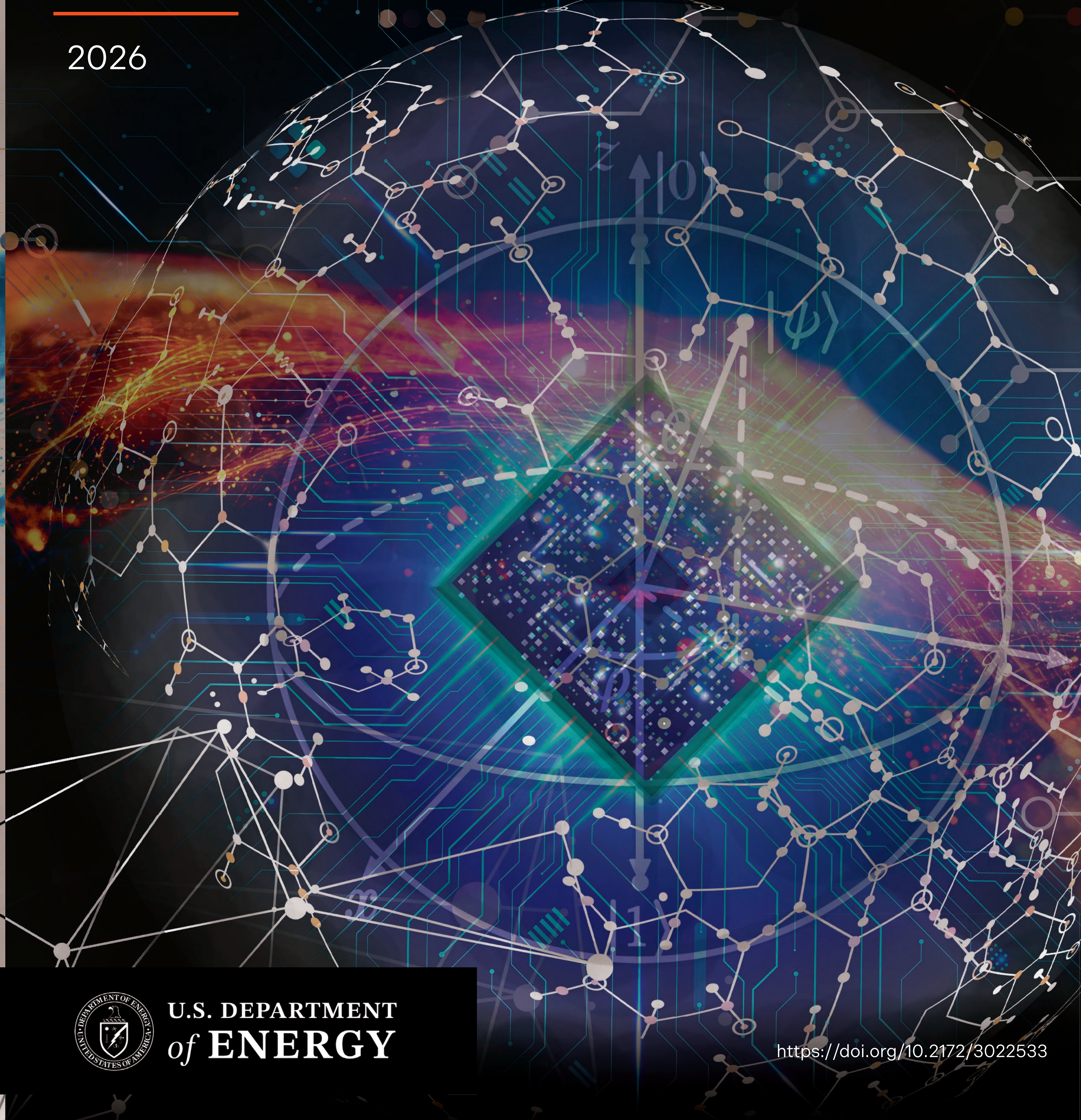

U.S. DEPARTMENT of ENERGY

# Blueprint for DOE Quantum Supercomputing

*Ensuring U.S. Leadership in the Quantum Decade*

**Committee and Report Authors:**
DOE Laboratory Quantum Supercomputing Blueprint Team: Ryan Bennink[1], Robin Blume-Kohout[2], Salman Habib[3], Andrew Landahl[2], Hai Ah Nam[4], Laura Schulz[3], Mallikarjun (Arjun) Shankar[1], Nicholas Wright[11]
DOE Points of Contact: Tanner Crowder[12], Pavel Lougovski[12]

**Significant Contributors**:
Brian Austin[4], Andrew Baczewski[2], Christian Bauer[4], Thomas Beck[1], Tanmoy Bhattacharya[5], Daan Camps[4], Zohreh Davoudi[7], Al Geist[1], Natalie Klco[8], Katie Klymko[4], Hank Lamm[6], Thomas Maier[1], Mason Rhodes[9,2], Martin Savage[10]

[1]Oak Ridge National Laboratory
[2]Sandia National Laboratories
[3]Argonne National Laboratory
[4]Lawrence Berkeley National Laboratory
[5]Los Alamos National Laboratory
[6]Fermi National Accelerator Laboratory
[7]University of Maryland
[8]Duke University
[9]University of New Mexico
[10]University of Washington
[11]NVIDIA; work performed when author was at LBNL
[12]Department of Energy

Cover image credit: Cathy Jenkins/Oak Ridge National Laboratory

# Contents

# Executive Summary

Quantum computing stands at the threshold of a transformative decade, where the field will evolve from small-scale demonstrations toward practical scientific computing at scale. This Blueprint identifies fault-tolerant quantum computers (FTQCs) as a viable, scalable, and broadly applicable path to achieving “quantum scientific utility,” defined as solving scientifically valuable problems beyond the reach of conventional, *classical* computers. This capability is expected to show scientific demonstrations in the late 2020s and to mature in the early-to-mid 2030s. This Blueprint outlines a strategy to prepare the U.S. Department of Energy (DOE) for FTQCs and their integration into the U.S. national scientific computing infrastructure. Its purpose is to identify the steps, milestones, and research directions necessary for DOE to enable initial deployment of FTQCs in 2028 as a scientific tool for the nation and mature this capability into the 2030s.

DOE has a long history of supporting quantum information science and technology, contributing significantly to research advancements, training a quantum-ready workforce, and providing access to early small-scale quantum hardware. Given recent demonstrations of logical operations on error-corrected logical qubits and the advancement of commercial hardware roadmaps, DOE should begin preparations for large-scale, fault-tolerant quantum computing deployment for DOE science missions.

This Blueprint proposes that DOE focus on (1) deploying first-generation scientifically relevant quantum computers with at least 100 logical qubits and performing at least 10,000 to 100,000 hard logical operations[1] in scientifically relevant calculations; (2) developing essential FTQC programming competencies, system software, and facility readiness; and (3) investing in cutting-edge focused R&D that fosters breakthroughs in scientific applications, algorithms, and logical architectures needed to accelerate the advent of scientific utility. This effort will position DOE to transition to larger systems: production-scale quantum computers that comprise 1,000 to 10,000 logical qubits, perform 1 to 10 billion hard logical operations, and execute scientifically useful computations at scale.

Achieving these goals will require DOE facilities to evolve with urgency to support scientific campaigns that integrate quantum and classical computing resources into efficient workflows, novel software and firmware environments for compiling and routing quantum programs on FTQC machines, and suitable infrastructure for quantum hardware. It will also require further development and optimization of scientific applications from the fields of materials science, quantum chemistry, and high-energy and nuclear physics. The Blueprint calls for transformative R&D and collective action to accelerate the advent of scientific quantum utility and bring it within reach by 2028.

[1] Throughout this document, performing a certain number of logical operations implies that they are performed with tolerably small, accumulated error.

# 1. Introduction

To position the U.S. Department of Energy (DOE) for global leadership in quantum computing for scientific discovery, it is important to urgently develop and deploy a mission-driven quantum computing capability and ecosystem. This initiative requires defining and aligning DOE applications with co-designed computer architecture, software, facilities, and resource roadmaps that lay the foundation now for initial capability deployment by 2028.

Quantum computation promises the ability to solve certain computational problems—for example, simulating and calculating properties of native quantum physical systems such as molecules, materials, and quantum fields—exponentially faster than any conventional computer. Harnessing this transformative capability aligns with DOE's mission to lead the world in scientific discovery. For decades, DOE and its 17 national laboratories have incubated, created, hosted, and used the world's most capable supercomputers and facilities to push scientific frontiers and solve science problems beyond the reach of pure theory and observational experiments. Building on its global leadership in deploying numerous first-of-their-kind supercomputers, DOE is poised to deliver era-defining quantum supercomputing capabilities. This Blueprint presents the motivation, the justification, and a high-level plan for DOE to drive an effort to establish quantum computing capabilities in concert with its existing support for strategic quantum computing R&D. Scientifically relevant quantum computers (SRQCs) will be unique in their requirements for housing, operation, and programming. It is critical for DOE to prepare for their advent now with the following actions:

1. Accelerate national competencies for SRQCs and production-scale quantum computers (PSQCs), analogous to the United States' world-leading supercomputing ecosystem:
   a. Create expertise in mapping scientific applications to SRQCs and co-design SRQC architectures to implement them; grow expertise and the nation's workforce more generally to support and use SRQCs.
   b. Invest in toolchains to automate large-scale compilation of algorithms to fault-tolerant quantum computers (FTQCs).
   c. Develop standardized assessments and benchmark frameworks for FTQCs.
2. By 2028, deploy SRQCs that implement a complete and programmable suite of fault-tolerant logical operations; enable at least $10^4$ to $10^5$ hard logical operations—such as $T$ or Toffoli gates—on at least 100 logical qubits to validate the promise and viability of FTQCs; and demonstrate scaling[2].

This set of priorities and the accompanying timeline are consistent with projections for improvement in best-in-class quantum algorithms and logical architectures and with roadmaps of leading quantum computing hardware companies.

Figure 1 illustrates how goal-oriented investments in applications, algorithms, and architectures will enable DOE to lead in the path to a PSQC. The DOE complex is uniquely positioned with a critical mass (one that can be expanded) of the requisite quantum computing and domain science subject-matter expertise and leadership in scientific facility operations. Targeted investments to apply DOE expertise to SRQC co-design, acquisition, deployment, and use will dramatically accelerate progress toward DOE's desired capability goal.

[2] See the glossary for technical definitions of these terms.

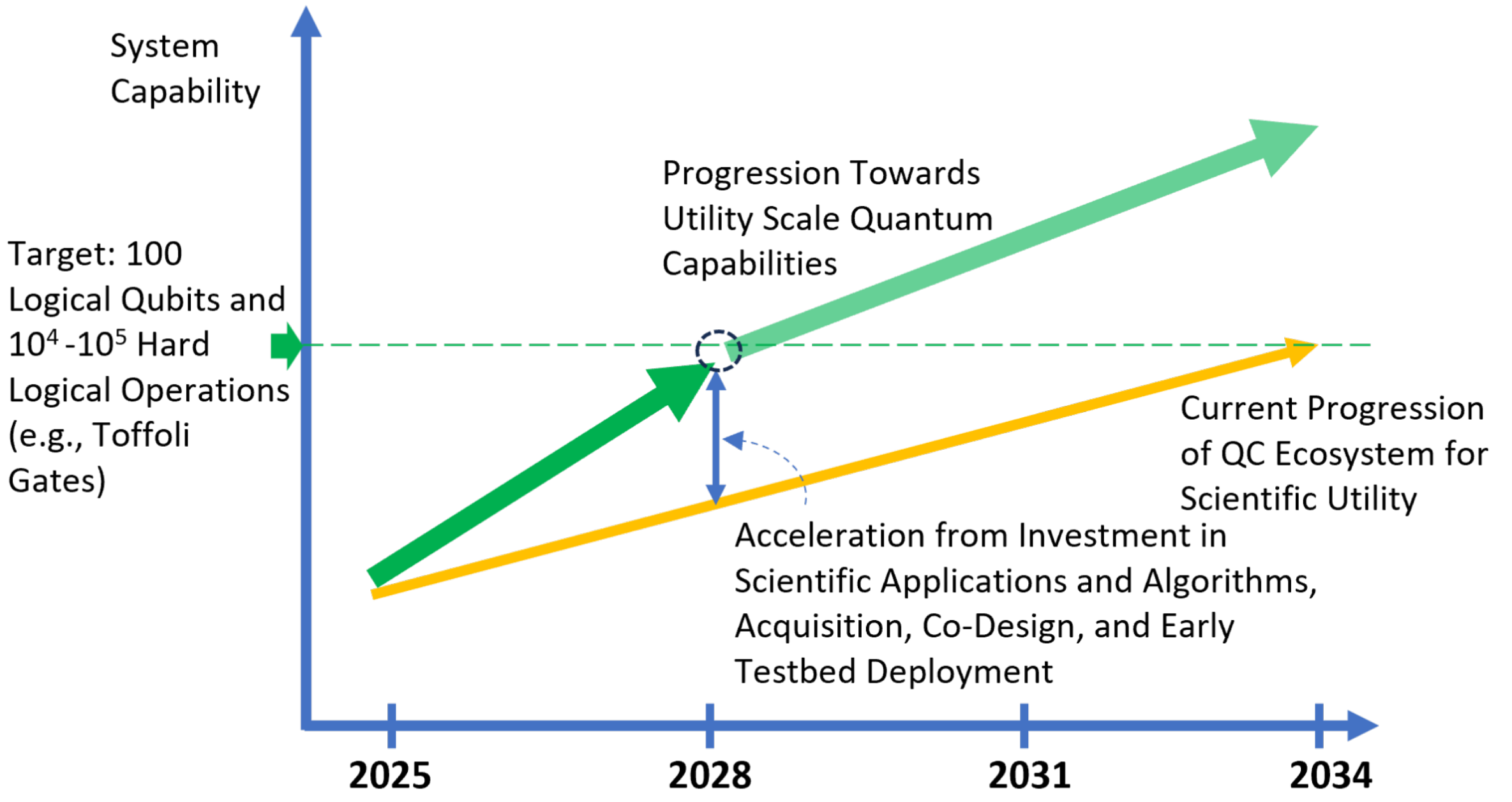

**Figure 1. Investments accelerate progress towards scientific utility of PSQCs.**

In the following sections, we discuss the landscape of FTQCs (Section 2), their application to representative DOE-relevant scientific missions (Section 3), FTQC capability roadmaps (Section 4), systems software and facility considerations (Section 5), acceleration through goal-driven R&D (Section 6), and criteria for success (Section 7). We offer a suggested roadmap for next steps (Section 8). Finally, the Appendix to this report contains an outbrief from a research roundtable meeting organized by the national laboratory authors of this report; the Appendix covers essential R&D activities the community needs to undertake for the roadmap outlined in Section 8 to be successful.

# 2. The Landscape of Quantum Computation

> The most promising quantum computing paradigm for scientific problems is fault-tolerant quantum computing, which processes logical qubits encoded into quantum error correction (QEC) codes. This will be implemented on special-purpose computing machines that are expected to coexist and cooperate with leadership-class high-performance computing (HPC) systems on scientific campaigns.

A wide range of paradigms for SRQCs have been proposed. In the noisy intermediate-scale quantum (NISQ) era, non-fault-tolerant paradigms such as adiabatic evolution, annealing, analog simulation, variational computing, and bare physical computation were considered. Although several of these paradigms have the potential to enable some novel domain-science discoveries [Kim2023, Alam2025a, Alam2025b], FTQCs are the most promising approach for providing scalable solutions to science problems in multiple domains that are well beyond the reach of classical supercomputers [DOE2024]. This report therefore focuses on the prospect of FTQC architectures (also called *logical architectures*) and machines and on the algorithms that run efficiently on them.

## 2.1 Fault-Tolerant Quantum Computers for Science

FTQCs are a rapidly emerging technology. The first demonstrations of (tiny) end-to-end fault-tolerant computations, according to the rubric proposed by Gottesman almost 10 years ago [Gottesman2016], appeared in 2024–25 [Reichardt2024a, Reichardt2024b, Bluvstein2025]. More critically, all the key components that will enable large-scale FTQC technology—error-corrected logical qubits [Google2025a], transversal logical gates [Egan2021, Paetznick2024], and "hard" nontransversal logical operations [Postler2022, Dasu2025]—have been demonstrated. These point to the feasibility of increasingly large FTQC computations in the coming years. Extensive theory and design have led to roadmaps for how to develop hardware to a scale sufficient for PSQCs.

Two widely used metrics for an FTQC are the number of *logical qubits* it provides and the number of *hard logical operations* that it can perform before accumulating substantial computational errors. We provide estimates of how many logical qubits and hard logical operations are sufficient to perform the core computational tasks for several exemplar science challenges in Section 3. In Section 4, we assess the resource requirements for these given the projected capabilities of quantum hardware over the next 5+ years.

## 2.2 Synergy Between Quantum and Classical Supercomputers

As currently envisioned, quantum computing will be a specialized technology that will provide transformative advantages on specific classes of computational problems that impact DOE's mission. Scientific campaigns that resolve grand open challenges—such as "How does high-$T_c$ superconductivity work?", "What is the nature of transition-metal catalysts?", and "How do quarks and gluons dynamically form the matter we observe?"—with the help of quantum computation will also rely heavily on conventional HPC, a powerful resource in its own right. This report is motivated by a vision of quantum and classical supercomputers coexisting and working together on integrated scientific campaigns.

As large FTQCs become available, at least three modes of joint operation with HPC resources are conceivable:

- *Independent*: A quantum algorithm is executed on an FTQC. Its inputs and/or outputs may be pre- or postprocessed by an HPC system offline.

- *Loosely coupled*: A classical control program delegates subroutines to FTQC and HPC resources, whose inputs are defined by conditional logic using results of prior tasks. The FTQC-HPC interfaces occur at relatively low latency/bandwidths. FTQC and HPC resources may be accessible through a common API and may only require latency/bandwidth consistent with cloud access.

- *Tightly coupled*: Hybrid algorithms run simultaneously on FTQC-HPC resources, interchanging data at latencies/bandwidths comparable to heterogeneous CPU-GPU systems and motivating specialized interconnects and APIs.

While independent and loosely coupled approaches are better represented in the current literature, tightly coupled hybrid algorithms remain under active investigation. Useful FTQC subroutines are expected to require many (>10,000) logical operation clock cycles at slow logical operation clock speeds (<100 kHz) to yield minimal data (10 to 1,000 bits). Tight coupling may become more important in future FTQC architectures if FTQC algorithmic output data bandwidths increase.

It is worth emphasizing that below the algorithmic level, an FTQC is universally expected to require very significant embedded classical computing resources—as much as 50 petaflops at 1 Tb/s bandwidth and 20 μs latency [Camps2024, Caldwell2025]—to decode error-correction syndromes; run the logical microarchitecture; manage layout, scheduling, and routing in a way that is adapted for runtime failures; and perform on-the-fly calibrations. Integrating tightly coupled computational resources in the control plane to implement FTQCs represents a significant development challenge.

# 3. Solving Domain-Science Problems on Fault-Tolerant Quantum Computers

Early demonstrations of quantum scientific utility will likely emerge in at least three DOE-relevant domains: materials science, quantum chemistry, and high-energy and nuclear physics (HENP). In each domain, critical scientific questions rely on simulations of complex quantum systems—simulations that are beyond the limits of even exascale classical computation. Although current analyses suggest that FTQCs with 1,000+ logical qubits and the ability to perform billions of hard logical operations might be required to outperform cutting-edge HPCs, focused research into algorithms, applications, and FTQC architectures could enable scientific campaigns to be launched with on the order of 100 logical qubits and hundreds of thousands of hard operations.

A significant portion of DOE HPC workloads is devoted to three domains—materials science, quantum chemistry, and HENP—for which quantum computing offers a compelling promise of benefit [Camps2025]. Many urgent questions in these domains can be related to the properties or behaviors of an underlying complex quantum system. Efficient and accurate simulation of quantum physical systems on quantum computers [Feynman1982, Lloyd1996] is arguably the best near-term prospect for advancing these domains beyond the reach of post-exascale HPC.

This section presents three representative use cases for quantum computing that span these three domains, and gauge the minimum size and quality of a quantum computer needed to demonstrate scientific utility. We have selected these problems because they (1) have significant scientific value, (2) are beyond the reach of exascale-class HPC, and (3) are among those likely to be within the reach of FTQCs the soonest. Each discussed problem represents a broad class of scientifically valuable problems [Bauer2020, Bellonzi2024, Alexeev2024, DiMeglio2024] that could all be solved by an FTQC of comparable size. The FTQC requirements we present for these problems are based on published analyses of state-of-the-art algorithms. However, quantum algorithms are constantly improving, and the "first fruits" of quantum computing are a moving target [Proctor2025].

## 3.1 Assessing Prospects for Utility

*Resource estimation* [Harrigan2024, Gheorghiu2025] involves a detailed investigation of how an algorithm for a specific problem could be compiled and implemented on a specific proposed FTQC architecture and, critically, what expensive resources it would require. Resource estimates for applications based on worked-out implementations of established algorithms and architectures provide approximate upper bounds (worst-case estimates) for the quantum computational cost of a problem. Over the past 30 years, innovations in every part of the quantum software "stack," including algorithms, compilation, routing, and the underlying logical architecture, have contributed to a steady and dramatic reduction in the estimated resource costs of flagship computational problems. We expect this trend to continue, especially with focused investments by the DOE. Today, FTQC resource estimates exist for many scientific problems. Some are believed to be extremely conservative and thus likely candidates for significant improvements through continued research [Dalzell2023]. Others, especially those for canonical "challenge problems" expected to represent the earliest attainable examples of quantum scientific utility [Proctor2025], are the result of extensive research and may be difficult to reduce in complexity by more than 10×–100×. Importantly, every known application admits space-time (qubit-runtime) trade-offs that allow algorithms to be squeezed into fewer logical qubits (space) at the cost of more logical operations (time) or vice versa [Fowler2012b].

## 3.2 Materials Science

### 3.2.1 Description

Quantum materials are widely regarded as foundational for driving transformative advances in energy-relevant technologies across microelectronics, information, and computing systems. A key goal of modern physics is to develop a detailed understanding of the microscopic mechanisms that give rise to emergent and exotic quantum states exhibiting magnetic, superconducting, and topological order. The (Fermi-)Hubbard model is one of the simplest models that exhibits, even at small scales, some of the phenomena observed in the high-temperature superconducting cuprates, including Mott insulating, antiferromagnetic, strange metallic, and superconducting states. However, classical supercomputers cannot simulate the Hubbard model at sufficient size and energy accuracy to address grand challenges such as high-temperature superconductivity. Simulation of the Hubbard model is therefore a promising, significant application for quantum computing. Furthermore, it is characteristic of a broad class of valuable problems in condensed matter physics.

### 3.2.2 Candidate Workflow

A promising approach to solving lattice models such as the Hubbard model is to treat a large material system as a small cluster embedded within a large bath. The cluster is solved accurately, including all the complex many-electron interactions, while the bath is treated approximately (e.g., using mean-field theory). This suggests a loosely coupled quantum-classical workflow (Figure 2) in which the quantum computer simulates the cluster depending on the bath properties and the classical computer simulates the bath depending on the cluster properties [Bauer2016,

Vorwerk2022]. Quantum embedding approaches can also be combined with other approximations, such as density functional theory, to yield fully *ab initio* materials calculations.

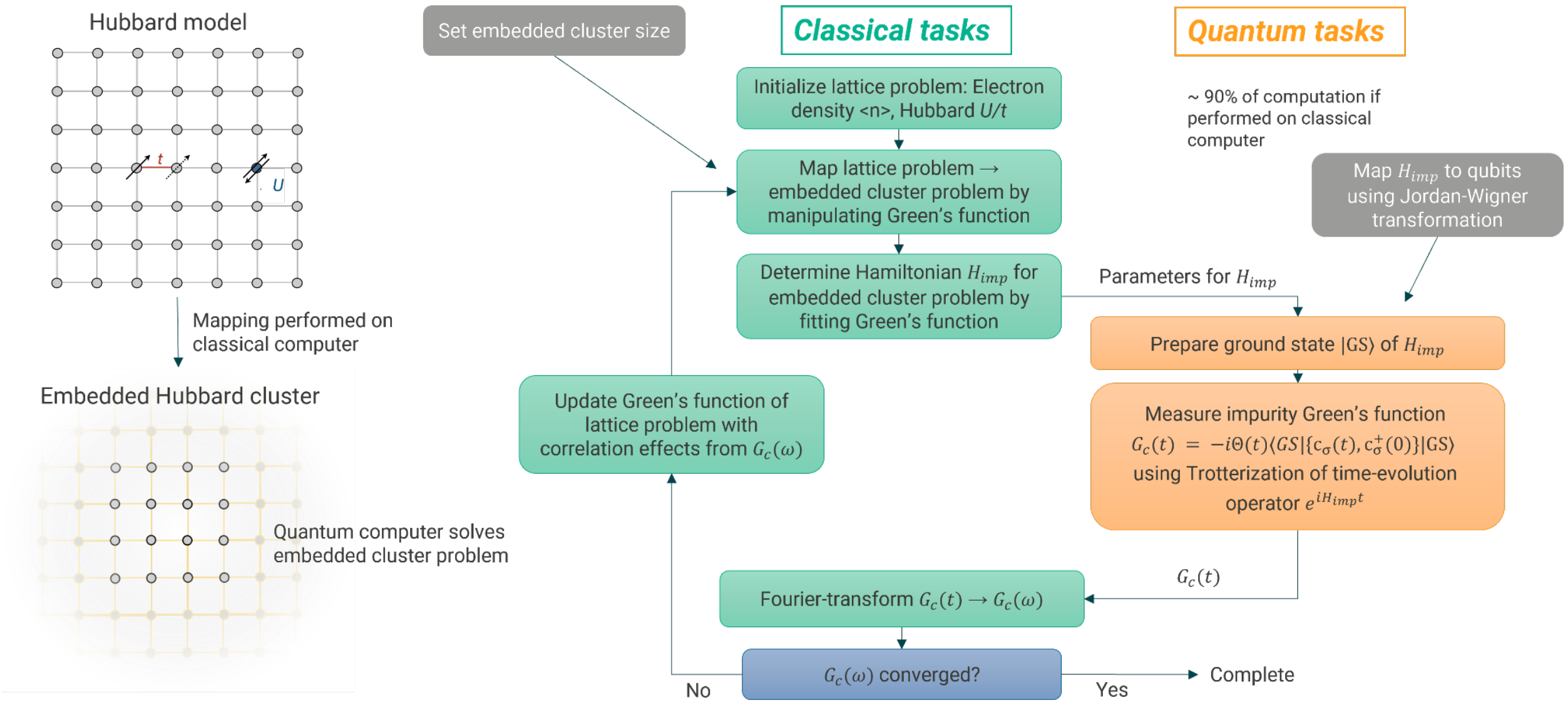

**Figure 2. Quantum simulation of embedded Hubbard model.** Starting with a nominal single-particle Green's function, a classical computer determines an effective cluster Hamiltonian and approximate ground state. On a quantum computer, the approximate ground state is prepared, refined, propagated forward in time according to the effective Hamiltonian, and finally measured in order to estimate a single-particle excitation-deexcitation amplitude. From a series of such simulations, each repeated many times, the time-dependent single-particle Green's function is estimated more accurately. This estimate is used to inform the next iteration, and the loop proceeds until the estimated Green's function converges, indicating a self-consistent solution.

### 3.2.3 Target Instances and Solution Requirements

Recent proof-of-principle demonstrations of the hybrid workflow described above on NISQ computers have targeted problems consisting of a single impurity site coupled to up to three [Keen2020, Steckmann2023, Nie2024, Hogan2025] bath sites. The accuracy of the cluster embedding model generally increases with both cluster size and bath size. A four-site Hubbard model cluster with eight bath sites is at the limit of classical computing capabilities. Solving a $6 \times 6$ embedded cluster model would lead to new scientific results that could potentially be validated by experiments on twisted bilayers of transition metal dichalcogenides, which are believed to be good realizations of the Hubbard model. Resource estimates for quantum phase estimation (QPE)—a related computation that may be expected to have comparable costs—suggest that simulation of such a cluster could be accomplished with 200–300 logical qubits and $10^8$ hard logical operations [Campbell2021, Yoshioka2024, BaySmidt2025].

## 3.3 Quantum Chemistry

### 3.3.1 Description

The field of quantum chemistry presents a vast range of important scientific and technological challenges that offer vast potential for commercial and scientific utility. Several of these appear uniquely amenable to analysis on quantum computers, including efficient catalysts for fertilizers (nitrogen fixation); the discovery of new drugs; and the development of advanced materials for batteries, photovoltaics, and quantum information technologies.

Chemists simulating these problems on classical computers are forced to choose between accuracy and efficiency in their calculations. Some computational methods are efficient but vulnerable to systematic uncontrolled approximation errors, while others can achieve arbitrarily good accuracy at the cost of computational infeasibility (inefficiency). Quantum simulation algorithms offer the prospect of circumventing this dilemma, enabling accurate and efficient calculations when executed on sufficiently powerful quantum computers [McArdle2020, Babbush2023].

It is neither possible nor sensible to place an upper bound on the number of chemical systems for which a quantum computer could enable accurate and efficient calculations, as chemistry (and materials science) offer countless open problems because elements from the periodic table can be combined in combinatorially many ways. However, a few representative canonical problems have emerged in the quantum computing literature as clear exemplars of the potential impact of FTQCs. The best-known examples are FeMoco [Reiher2017, Low2025] and cytochrome P450 [Goings2022]. They represent important use cases for nitrogen fixation and drug metabolism, respectively.

Strongly interacting electrons are also the key to plasma physics, especially in the context of fusion energy [Joseph2023], a key DOE mission area where experimental data is rare and expensive, and attributing real-world behavior to atomic-scale processes is computationally hard. High-accuracy simulations of chemical processes in plasma-facing components could make magnetic fusion devices more reliable. Quantum simulations of degenerate plasmas can improve our microscopic understanding of materials for inertial fusion devices in experimentally inaccessible thermodynamic regimes where performance-limiting instabilities form [Rubin2024]. Quantum computations, assisted by AI and HPC, could yield more accurate models of solvation chemistry in molten-salt breeding blankets, enhancing production of tritium fuel required for both inertial and magnetic fusion. Quantum algorithms for fusion/plasma applications are similar to those for chemistry and materials, except that thermal and nonequilibrium effects cannot be neglected. In the future, quantum algorithms for solving partial differential equations describing macroscale plasma models [Liu2021, Ameri2023] (with connections to chemistry [Kharazi2026]) may have even greater impact for fusion science.

### 3.3.2 Candidate Workflow

Wholesale replacement of classical electronic structure methods with quantum computers is unlikely for problems of chemical interest, at least until the advent of PSQC systems. Nevertheless, it is expected that integrating quantum computing with classical HPC techniques can enable computations on SRQC systems that provide scientific insights [Alexeev2025]. A workflow for such a hybrid computation is shown in Figure 3. The basic idea is very similar to that described in the previous section on materials. A classical computation is performed to determine the properties of the overall system and determine the size of the smaller fragment to be treated more accurately using the quantum computer. The results from the quantum computer are then combined with the classical ones to determine the property of scientific interest. A convergence criterion is used to determine whether more calculations using the quantum computer are necessary.

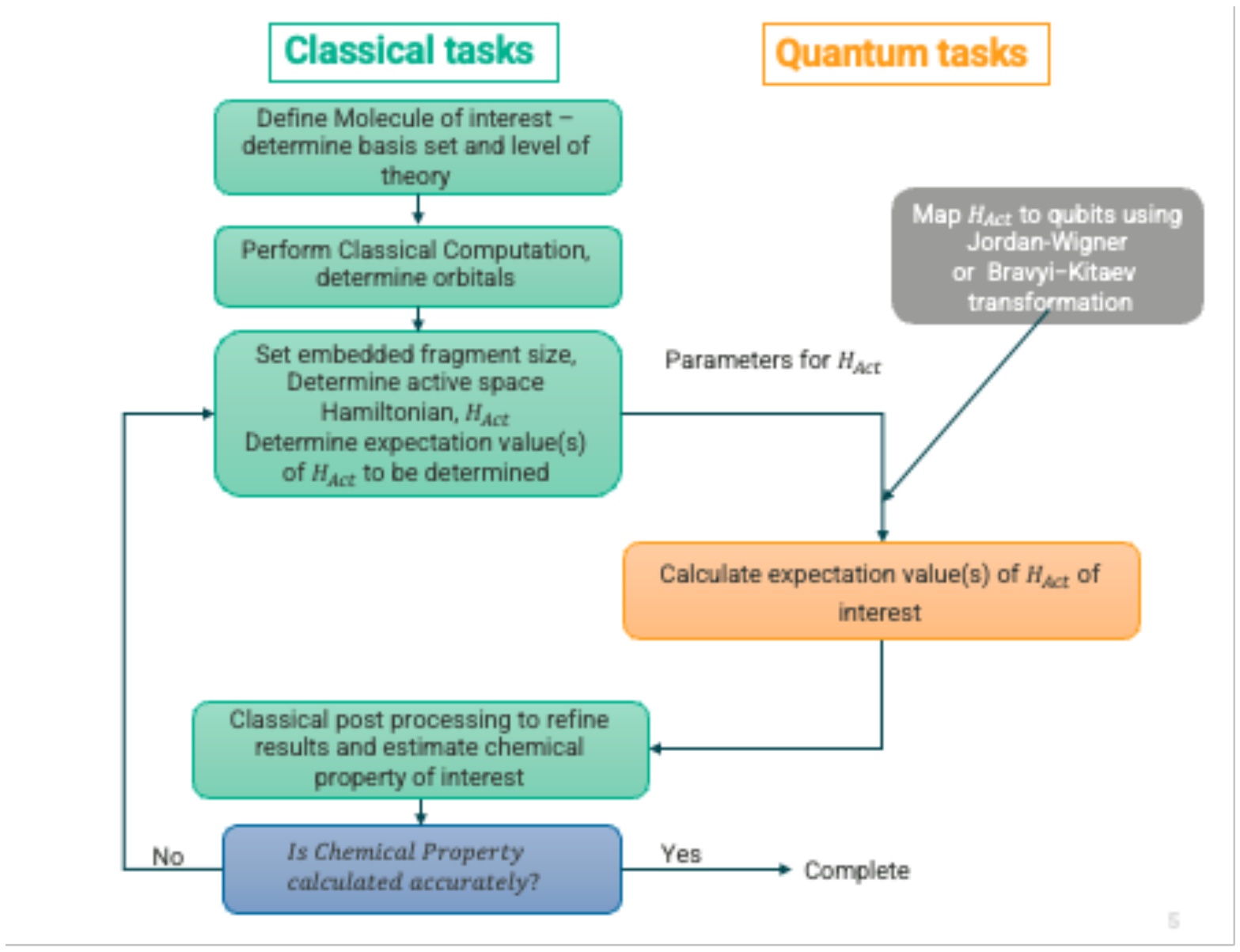

**Figure 3. Quantum simulation of a molecule with fragment-based treatment of the region of strong correlation [Alexeev2025].** First, the chemical property of interest, selected molecule, level of theory, and basis set are specified. An effective fragment Hamiltonian and approximate ground (or, if desired, excited) state are computed on a classical computer. An approximation to the target state is then prepared and refined on a quantum computer, and the expectation value of the Hamiltonian is estimated. This estimate is combined with prior information in a postprocessing step on the classical computer to yield a better estimate of the chemical property of interest, which is used to inform the next iteration. Iteration proceeds until the chemical property is obtained to the desired accuracy.

### 3.3.3 Target Instances and Solution Requirements

A recent report reviewing projected resource requirements for quantum chemistry systems suggested that solving scientifically relevant, hard quantum chemistry problems such as estimating the ground energy of FeMoco, using QPE, will require running at least $10^8$ hard logical operations on about 1,000 logical qubits, with estimates going up to $10^{12}$ or more hard logical operations, depending on the specific application and algorithm under consideration [Camps2025]. For problems such as those described in Figure 3, where insights from classical calculations are combined with those from a quantum computer, the requirements are much more modest: in the range of 50–100 logical qubits with $10^4$–$10^5$ hard logical operations for a QPE calculation of a ground- or excited-state energy [Alexeev2025].

## 3.4 High-Energy and Nuclear Physics

### 3.4.1 Description

The purpose of HENP is to discover new physics beyond our current knowledge by studying fundamental forces and particles, the nature of space and time, and the origin and evolution of the universe. Because HENP R&D is a core DOE mission that is not pursued by industry, in contrast to applied chemistry or material science, commercial vendors are unlikely to co-design quantum computers for applications in HENP without proactive input from DOE.

A primary challenge that quantum computing is expected to accelerate is performing calculations based on quantum chromodynamics (QCD), the theory of the strong force in nature. The strong force is responsible for binding quarks and gluons into hadrons, including protons and neutrons, the building blocks of matter, from atomic nuclei to neutron stars and supernovae. Conventional classical supercomputing has led to remarkable success in predicting static properties of hadrons from the underlying QCD interactions [Joó2019, Davoudi2021, Davoudi2022, Aoki2024]. However, the Monte Carlo sampling techniques that enable this success encounter a computational wall, namely sign problems, in addressing dynamical processes or systems at finite density: to converge to the answer, exponentially large resources are required as the system size increases. Even static properties are beyond computational reach for highly compact particles, such as some dark matter candidates, because the runtime of classical algorithms grows inversely with the eighth power of the particle size [DeTar2004, Farrar2017, Farrar2020].

Quantum computers offer a new way to execute first-principles QCD-based calculations of dense environments, nonequilibrium phenomena, and real-time processes—phenomena prevalent in the early universe, supernovae explosions, nuclear reactions, and high-energy collisions of particles and nuclei. The ability of quantum computers to naturally encode exponentially large Hilbert spaces and to enable important sign-problem-free algorithms rooted in Hamiltonian evolutions leads to expectations that such computers will have a decisive advantage over classical methods [Carlson2018, Beck2019, Asai2023, Beck2023, Bauer2023].

### 3.4.2 Candidate Workflow

Workflows for simulating many such physical systems of interest have commonalities. Led by HENP subject-matter experts (SMEs), one may start the quantum program by designing and performing suites of physics-aware digital quantum simulations that prepare an array of initial quantum states of interest, apply time-evolution operators derived from the Standard Model of particle physics (including low-energy effective field theories of QCD) to them, and furnish sets of physics-motivated measurements of the properties of these nontrivial states throughout their evolution. Using physics-aware algorithmic optimizations, aided by quantum algorithm SMEs, one formulates optimized subroutines that boost the performance of the state-preparation, time-evolution, and observable measurement stages within the accuracy requirements. Next, in conjunction with quantum architecture SMEs, one then maps and compiles the algorithm to hardware in an optimized way using the prevailing logical (fault-tolerant) quantum instruction set architecture (ISA). In collaboration with quantum hardware SMEs, one then executes this compiled algorithm on the hardware and returns a suite of results that can be used for physics analysis by HENP SMEs and integrated into science domain research programs. In the long term, the roles of the quantum algorithm, quantum architecture, and quantum hardware SMEs—currently AI assisted—may become more automated. In the near term and especially for SRQC machines expected in the 2028 time frame, their inputs remain critical because co-design is essential for deployment success. Additionally, novel mappings of gauge theories (e.g., Refs. [Illa2024, Kürkçüoglu2025, Siew2026, Than2025]) and AI-assisted hybrid computing methods (e.g., Refs. [Lin2024, Aktar2025]) can be prototyped. Figure 4 depicts this notional workflow.

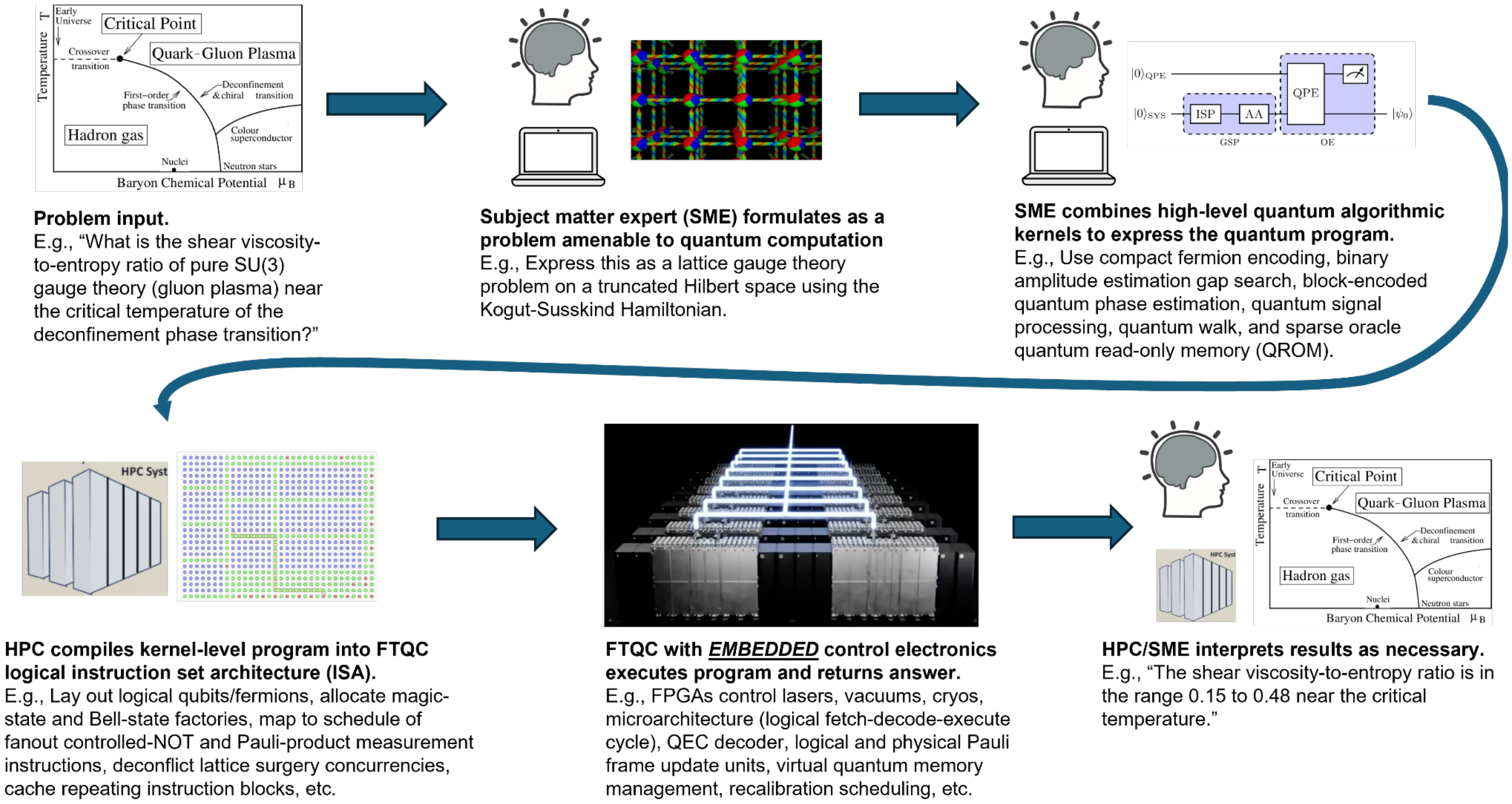

**Figure 4. Workflow for calculating the shear viscosity-to-entropy ratio of gluon plasma near the critical temperature of the deconfinement phase transition.**

### 3.4.3 Target Instances and Solution Requirements

As one example calculation, we consider digitized time evolution of a pure SU(3) lattice gauge theory [Kogut1975, Byrnes2006, Ciavarella2021, Ciavarella2024] (in the absence of dynamical fermions) to measure key observables, such as energy-energy correlations [Lee2025], distribution of particles and jets [Barata2022], and the viscosity and other transport coefficients of a finite-density medium [Cohen2021, Turro2024, Turro2025]. Perhaps surprisingly, pure SU(3) lattice gauge theory on just a $3 \times 3$ lattice in even two space dimensions and one time dimension is at the limits of what the most advanced supercomputers can simulate today. On the other hand, simulating a $10 \times 10$ lattice with a single-qubit truncation can be achieved with only 100 logical qubits and $3.2 \times 10^6$ $T$ gates over 50 Trotter steps. Modeling more interactions can require modest increases to $3.6 \times 10^6$ $T$ gates [Ciavarella2025]. A more faithful 5-qubit-per-plaquette representation increases the costs to 500 logical qubits and $4.7 \times 10^{11}$ $T$ gates [Ciavarella2025]. Complementarily, one can approach SU(3) via larger encodings using discrete subgroups [Assi2024] or anyonic regularizations [Rhodes2026]. On a smaller $3 \times 3$ system, the same simulation requires 144 logical qubits and $5 \times 10^7$ $T$ gates using an 8-qubit approximation [Perez2025], rising to 162 logical qubits and $4 \times 10^{10}$ $T$ gates with a 9-qubit approximation [Perez2025]. Progressing toward fully realistic simulations of QCD—which require three spatial dimensions, the inclusion of dynamical quarks, and increasingly precise approximations—will demand a larger number of logical qubits and lower error rates for hard logical operations [Kan2021, Rhodes2024]. Nonetheless, the algorithms and encodings required at these larger scales are direct extensions of those demonstrated on 100 logical qubits, ensuring that insights gained on early systems translate directly to the regimes necessary for full physical realism.

## 3.5 Common Algorithmic Motifs

The applications described above—like many applications expected to benefit from FTQCs—revolve around the simulation of a quantum physical system. Generally, a simulation algorithm takes a description of a quantum system's Hamiltonian and returns an estimate of a stationary or dynamical observable of that system. The ground-state energy is the most common stationary observable. Dynamical properties of interest include temporal correlation functions, phase transitions, and responses to external influences. Algorithms for simulating quantum systems on FTQC machines have been investigated since 1996 [Lloyd1996]. Although significant room for innovation remains, almost all simulation algorithms comprise three distinct subroutines:

1. Preparation of the qubits in a state that encodes an initial state of the physical system, often an approximation to the ground state.

2. Many (e.g., $10^4$) repeated applications of a unitary operator that encodes the Hamiltonian.

3. Measurement of a multiqubit operator that encodes a scientifically interesting observable.

Typically, this three-step sequence is repeated multiple times to obtain statistically accurate estimates of targeted observables.

Although the specifics of each subroutine differ across applications, a relatively small set of quantum computational kernels can be used to implement them all. An FTQC that can perform each of these kernels at scale will be able to run quantum algorithms for a wide variety of scientific applications.

A primary computational kernel is *Hamiltonian evolution*. This kernel applies $exp\ (-iHt)$ to a set of qubits, where $H$ is a Hermitian operator (often, but not always, the system Hamiltonian [Low2019]) and $t$ is the duration of evolution under $H$. The operator $H$ typically consists of a sum of relatively simple but noncommuting terms. Various techniques exist for approximately implementing short-time evolution with bounded error. Evolution for long times is implemented as a sequence of short-time evolutions, such as Trotter steps. Besides simulating dynamics, Hamiltonian evolution is also used in QPE to estimate energies.

The *quantum Fourier transform* performs a very efficient discrete Fourier transform on the vector of probability amplitudes that define the state of a qubit register. It is used, for example, within QPE to transform time-domain signals into frequency information.

QPE is a routine used to isolate eigenvalues and eigenstates of an operator. It is often employed to find the ground state and/or ground-state energy of a system's Hamiltonian. Traditional QPE applies a series of ancilla-controlled Hamiltonian evolutions to an initial state and then performs a quantum Fourier transform on the ancilla register. Reading out the ancillae yields an energy (eigenvalue) and leaves the system qubits in the corresponding eigenstate. The particular eigenstate obtained is generally nondeterministic. It depends on the energy distribution of the initial state; preparing a "good" (energy-localized) initial state can be harder than phase estimation itself.

*Amplitude amplification* is a method commonly used to boost the success probability of another quantum subroutine. It involves repeated application of that subroutine, interspersed with calls to two quantum "oracles" (black-box subroutines) that mark (i) the initial state and (ii) the target state or subspace whose amplitude is to be amplified. If the success probability of the original subroutine is $p$, amplitude amplification boosts the success probability to approximately 1 by calling the subroutine only $p^{-½}$ times (fewer than the $1/p$ calls necessary for classical probability amplification). Amplitude amplification is used to prepare initial states for phase estimation and for many other uses, including function inversion (or "search") and integration.

The *quantum singular value transform* is a meta-algorithm that unifies and generalizes many of the kernels above. Given a function $f$ and the ability to implement an operator $A$, it provides an efficient way to implement $f(A)$. This is a general and powerful technique that can be used for many tasks, including Hamiltonian evolution, imaginary-time evolution (simulated cooling), inversion, and spectral filtering.

Many other quantum algorithms and kernels are used within end-to-end quantum applications. The performance and resource requirements of the kernels above have been analyzed in a host of different contexts, leading to plausible cost and performance projections for many quantum tasks. Although the discovery of new quantum kernels has been historically infrequent, research in quantum algorithms continues to yield steady improvements in algorithmic efficiency.

## 3.6 Capability Requirements for Achieving Quantum Scientific Utility

This survey of applications indicates that scientifically significant quantum computations involve roughly a billion or more hard logical operations on several hundred or more logical qubits. Proposed FTQC architectures use 100–1,000 elementary quantum operations, or *quops* [Gidney2021], per hard logical operation, as well as additional logical qubits for data movement and implementation of hard operations. Thus, some current estimates for broad scientific utility require a *teraquop* machine that can accurately perform $10^{12}$ quops on 1,000+ logical qubits. We believe and expect that advances in algorithms and architectures, outlined in Section 6, will bring this estimate down over the next 4–8 years.

# 4. Capability Roadmap and Projections for Fault-Tolerant Quantum Computers

> Achieving general scientific utility will require FTQCs that greatly exceed today's capabilities. Today, ~5,000 operations can be performed on ~100 physical qubits. Roadmaps suggest the availability of megaquop quantum computers using advanced architectures that can perform 10,000–100,000 hard logical operations on 100 logical qubits circa 2028–29, followed by teraquop systems in the early 2030s. These milestones define a clear path forward and will require fault-tolerant architectures using QEC.

State-of-the-art quantum computers based on superconducting transmons, trapped ions, and neutral atoms can execute ~5,000 operations on ~100 physical qubits. They have demonstrated QEC [Egan2021, Google2025a, Bluvstein2025], quantum simulation of Ising [Kim2023] and Hubbard [Alam2025a, Alam2025b] models, and other accomplishments [Arute2019]. However, existing machines are orders of magnitude less performant than necessary to execute scientifically relevant computations that require $10^5$ to $10^9$ hard logical operations on 100 to 1,000 logical qubits.

The performance of such machines is often quantified by the error rate per *elementary quantum operation* (quop). However, algorithms are usually specified in terms of *hard* logical operations (e.g., $T$ or Toffoli gates), and each hard logical operation typically requires 100–1,000 quops. Thus, an FTQC that can execute $10^9$ hard logical operations is a *teraquop* (not *gigaquop*) machine. Logical architectures proposed for achieving teraquop performance use as many as 584 physical qubits per logical qubit [Gidney2025] or as few as 72 [Yoder2025], with extremely sensitive dependence on the quality of the physical qubits. Although improvement can be expected, the goal of $10^9$ hard logical operations on 1,000+ logical qubits is likely to require at least 100,000 high-fidelity physical qubits.

Quantum computers of rapidly increasing size and scale are being demonstrated by U.S. commercial companies. (Academic groups, U.S. national labs, and non-U.S. companies have demonstrated impressive but less capable systems.) Several industrial roadmaps are published and indicate a consistent trend over the next decade. Progress by any *individual* quantum hardware producer is unpredictable, but we judge it plausible that at least one platform will, at any given time, be able to produce a machine consistent with the following expectations:

- 2026: 1–10 logical qubits, $10^4$ quops
- 2027: 5–50 logical qubits, $10^3$–$10^4$ hard logical operations
- 2028: 25–100 logical qubits, $10^4$–$10^5$ hard logical operations
- 2029: 60–200 logical qubits, $10^5$–$10^7$ hard logical operations
- 2031–33: 1,000–2,000 logical qubits, $10^9$–$10^{12}$ hard logical operations

These moderately optimistic best-guess estimates are consistent with Figure 5 from a 2025 National Energy Research Scientific Computing Center (NERSC) study [Camps2025] that compared capability requirements for quantum scientific utility to vendor roadmap predictions for 2025, 2030, and 2035.

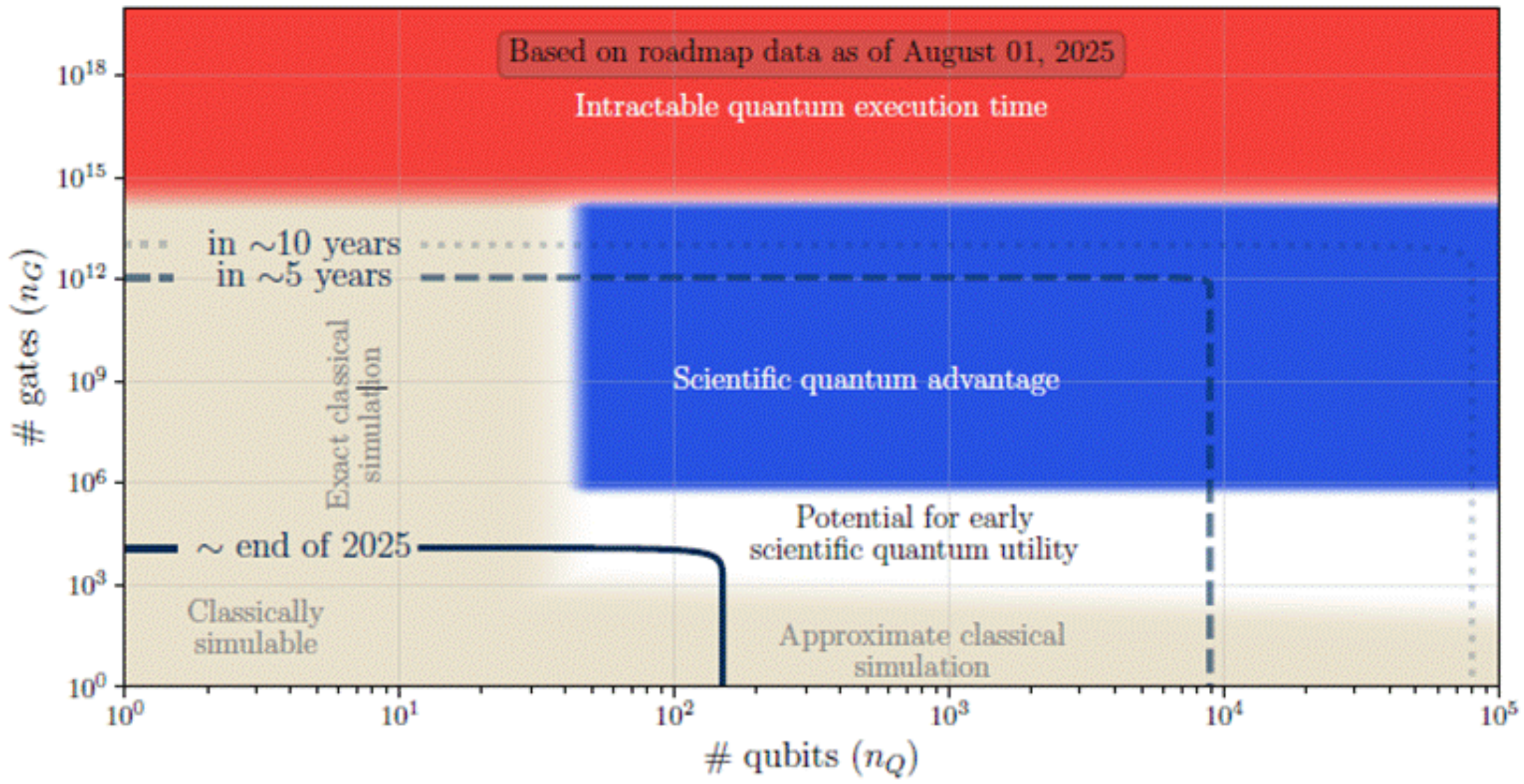

**Figure 5. Quantum hardware requirements for scientific problems [Camps2025].**

Barring transformative breakthroughs, best estimates from quantum algorithms (Section 3) and vendor roadmap projections suggest the following:

- Multi-megaquop FTQCs offering ~100 logical qubits and $10^4$–$10^5$ hard logical operations may become available at the end of this decade. These first-generation FTQCs will enable testing of the principles, scalability, logical ISAs, and software stack of FTQCs. They will function as scientific instruments, allowing for increasingly complex scientific calculations, and will demonstrate scaling paths toward PSQCs in the early 2030s.

- A teraquop FTQC may become available in the early 2030s. Such a PSQC would increase the capability and capacity of scientific campaigns.

These projections motivate priorities for achieving scientifically relevant quantum computing as rapidly as possible by pursuing three activities urgently and simultaneously:

1. **Co-design with vendors** to deploy a quantum computer that can perform at least $10^4$–$10^5$ hard logical operations on ~100 logical qubits targeting the 2028 time frame.

2. **Begin preparing facilities, user bases, expert operators, and software/theory infrastructure** necessary to make immediate, effective use of such a quantum computer by 2028.

3. **Scope and fund urgent, breakthrough R&D** focused on achieving transformative leaps that could (individually or collectively) enable quantum scientific utility to be achieved *early* on an FTQC.

# 5. Software, Systems, and Facilities Preparations

Over the next five years, DOE should fully develop a system software environment that makes SRQC accessible to a range of users as a managed resource. DOE should adopt and/or create logical instruction schedulers, compiling/routing software, and intermediate representations that facilitate the flexible use of diverse FTQC architectures across all application regimes without assuming a particular underlying quantum computing technology.

By 2028, to successfully host the first SRQCs, DOE facilities should evolve from experimental hosts to operational environments for quantum systems through investments in facility infrastructure, environmental control, and integrated management. These facility approaches may vary depending on the requirements of specific FTQC technologies and/or deployment models. Establishing unified operational procedures, secure access frameworks, and vendor coordination will make quantum computers reliable, serviceable components of the national supercomputing infrastructure.

DOE should lead the integration of quantum computing into the national scientific infrastructure, leveraging its unique scale, coordination capacity, and long-term stewardship across the U.S. national laboratory complex. Building on its existing ecosystem of supercomputers, specialized AI systems, high-throughput data platforms, and large scientific instruments, DOE should develop a comprehensive software and systems environment that will allow FTQCs to be managed as production-class resources. DOE should complete foundational work in unifying hardware control, scheduling, compilation, and user-access interfaces inspired by similar HPC facility processes. When SRQC deployment models require coupling to HPC resources, the DOE should establish integrated approaches to HPC facility workflows, access models, and governance. This enables high-level scientific applications to delegate computations to HPC resources when SRQC capabilities are not required—for example, in pre- and post-processing of data sent to and received from SRQC machines. Community-driven development of schedulers, compilers, and intermediate representations can facilitate portability across architectures. The 2028 SRQC(s) will serve as the first proving grounds for scientific discovery and full-stack integration.

## 5.1 Quantum-Classical Interaction Layers

The optimal configuration of how classical computing can best enable emerging application and operational needs for FTQCs is an area of active research. In general, across a range of HPC-FTQC coupling scenarios, the following three layers of classical computing could enable a scientific campaign running with and around an FTQC system:

1. **QEC and Fault-Tolerance Control Layer:** This layer provides a real-time (low latency and high-bandwidth state processing) exchange between classical compute and quantum hardware, performing continuous data processing, error decoding, and device configuration to maintain overall system stability. It is expected to be procured with the quantum system and may employ application-specific integrated circuits (ASICs) or field-programmable gate arrays (FPGAs) to manage I/O operations and low-level hardware setup specific to the underlying qubit technology. To achieve error correction or fault tolerance, more than one computational sub-layer may be needed. To enable the DOE to make best use of this layer, it would be valuable for the DOE to co-design the QEC and FTQC interfaces for applications it is interested in, much as it has historically done in conventional HPCs. For example, error-corrected logical fermionic data types and methods for processing them fault-tolerantly could support the exemplar applications listed in Section 3.

2. **Quantum-Classical Orchestration Layer:** This layer controls execution by utilizing classical computing capabilities to manage compilation, scheduling, and coordination of SRQC resources, linking user workflows to the hybrid compute environment (Figure 6). In some cases, this tier may be procured with the quantum system; in others, it may reside within a larger HPC complex. In all cases, it must include sufficient compute resources and libraries to perform compilation, scheduling and hybrid resource management, telemetry aggregation and response, and overall orchestration of the job workflow through the SRQC system. This will become especially important as we scale the number of logical qubits up, because algorithms for optimal compilation often demand computational resources that scale as high-degree polynomials as a function of the number of qubits involved.

3. **Scientific Ecosystem Workflow Layer:** The quantum supercomputers procured and deployed by DOE will be a part of the broader DOE Genesis Mission's infrastructure. This science-facing infrastructure will support end-to-end research workflows and will deliver cutting-edge scientific outcomes. It will execute the classical components of scientific discovery workflows, encompassing data preprocessing, model calibration, and post-quantum analysis. This layer connects quantum execution outputs with large-scale HPC and AI resources for data assimilation, parameter optimization, and uncertainty quantification. The Genesis Mission infrastructure is expected to enable integration with DOE user facilities and experimental instruments, allowing direct ingestion of observational or experimental data to drive workflows encompassing hybrid simulation, training, and validation loops.

The examples in Section 3 illustrate DOE scientific missions that integrate quantum and classical components and require advancements across all layers. The quantum portion executes on the FTQC subsystem, whose architecture, error rates, and runtimes may vary across technologies [Buchs2025] and circuit/system architectures [Bravyi2022]. A low-latency, high-bandwidth classical control layer is critical to sustain QEC and fault tolerance. Quantum-derived results can be exchanged with classical supercomputers within outer workflow loops that coordinate iterative quantum-HPC execution cycles.

## 5.2 Software Implementation

A software ecosystem for scalable quantum supercomputing should address the layers described in Section 5.1. Although a complete solution is several years away, ecosystem building efforts will be a part of the SRQC pathfinding and help bolster success in the PSQC phase.

The primary areas of R&D required to develop this coordinated software stack are:

1. **Hardware Abstraction and Device Management:** Although vendors may initially aim to develop their own hardware interfaces and device management technologies, it is in DOE's interest to co-design and jointly develop a cross-vendor solution and interface through R&D. This is essential for portability across vendors and would help to move the ecosystem toward standardized and consistent functions for initialization, calibration, execution, telemetry, connectivity, and error rates integrated into DOE's facility-monitoring and maintenance systems. Establishing the right level of standardization across different logical architectures will accelerate the deployment and adoption across the scientific user base.

2. **Compilation and Optimization Frameworks:** Development is needed in hybrid compiler pipelines to map high-level quantum algorithms into portable intermediate representations and dynamically optimize them for multiple back ends. Co-designing logical ISAs using innovations in codes, heterogeneous processing and memory elements, and other technologies would enable DOE to deploy logical ISAs for their applications onto fixed physical hardware with fixed physical ISAs developed by vendors. Advanced compilers will likely use telemetry and error-correction data to adapt scheduling and resource optimization on the fly, giving users visibility into achievable precision, depth, and runtimes. AI-assisted optimization could accelerate circuit mapping and performance tuning. Ongoing efforts such as the Munich Quantum Software Stack [Burgholzer2025], RIKEN's integration of quantum systems with the Fugaku supercomputer [Caldwell2025], and work led by Oak Ridge National Laboratory [Beck2024, Shehata2025, Shehata2026] demonstrate the value of effective modular design, open interfaces, and shared telemetry. DOE will build on this experience, particularly in portability, orchestration APIs, and standardized metrics, while leading efforts to develop an industry-wide solution that encompasses the needs of the DOE mission.

3. **User Access and Application Interfaces:** Unified portals, APIs, and workflow systems should be delivered through existing DOE supercomputing environments, where applicable. Support for standardized quantum frameworks and domain-specific libraries will allow researchers to launch hybrid jobs using familiar authentication, storage, and scheduling tools. These quantum frameworks should be designed so that they can utilize DOE's existing HPC infrastructure. For workflows that require only independent or loosely coupled HPC support, such integration is less imperative, whereas other workflows might rely on it more substantially.

The current vision of programming a quantum computer to solve science problems involves a hand-crafted sequence of steps as illustrated in Figure 6 (and earlier in the HENP application workflow illustrated in Figure 4). Mapping a class of scientific problems to an FTQC for the first time requires significant domain expertise at many of the stages indicated in these figures. AI may play a role in facilitating this innovation chain. Regardless, we expect substantial computing resources will be used in this endeavor, including HPC resources. Eventually, once a class of problems has been mapped, it is widely believed that much of the chain can be automated and optimized.

The execution flow illustrated in Figure 6 provides a framework for understanding how choices at different layers of the stack propagate through the system and ultimately determine whether a scientific application can be mapped efficiently onto available quantum and classical resources. At the bottom of the stack, different qubit technologies have vastly different characteristics in terms of number and connectivity of qubits and the speed and fidelity of physical operations. Within a given platform, multiple implementations of fault-tolerant protocols may exist, effectively representing distinct firmware variants with different choices of QEC code and/or logical instruction set architecture. The layout of a computation in terms of computational regions dedicated to memory, data movement, computation, and resource state production presents additional options. Finally, most quantum algorithms offer trade-offs between qubit count, number of operations, and precision of the quantum calculation. Some applications and algorithmic strategies even allow a trade-off between classical and quantum computation.

Collectively, these choices create a large and interdependent design space that can greatly impact the net resource requirements and performance of an application, making end-to-end analysis and orchestration across the software and systems stack essential. This requires an integrated software and systems stack spanning applications, algorithms, compilation and runtime systems, orchestration frameworks, facility operations, and the libraries and resource-estimation capabilities needed to effectively distribute and coordinate computations across classical and quantum resources.

As reflected in the lower stages of the Figure 6 cascade, major challenges remain in supporting the real-time QEC decoding, control, and data movement required for SRQC systems and beyond. Addressing these challenges will require advances not only in device technology, but also across the hierarchy of classical computing infrastructure supporting quantum systems, ranging from near-device control and decoding capabilities to higher-level networking, scheduling, orchestration, and runtime management for hybrid scientific workflows. Because QEC implementations, logical ISAs, compiler strategies, and execution models are tightly interrelated, these layers are natural targets for long-term co-design among DOE facilities, vendors, software developers, and application teams.

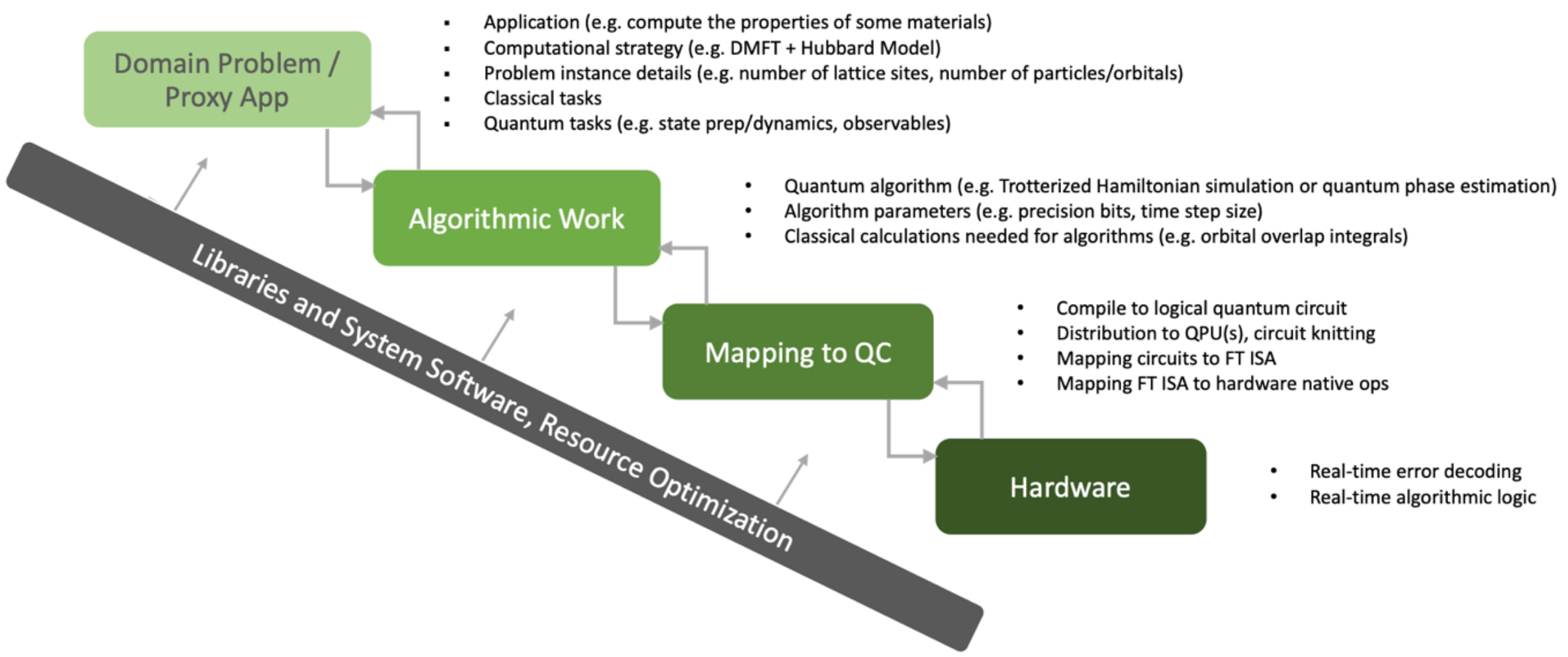

**Figure 6. Architectural workflow for mapping scientific applications onto fault-tolerant quantum computing systems through coordinated algorithm design, compilation, resource estimation, and execution management.**

Given the complexity and interdependence of these software, hardware, and operational layers, DOE should begin building an integrated operational ecosystem for quantum-enabled scientific computing during the early SRQC era. This includes developing the software and orchestration frameworks, governance structures, allocation models, accounting mechanisms, and facility policies needed to coordinate heterogeneous HPC, AI, and quantum resources across distributed and co-located environments. Establishing these early operational foundations will accelerate scientific application development, reduce ecosystem fragmentation, and help ensure DOE retains a globally competitive leadership position as quantum computing advances toward practical scientific utility.

## 5.3 Facilities Integration

Connecting quantum computers with DOE's HPC centers may require targeted preparation across facilities, infrastructure, and operations. Quantum systems can impose unique environmental, power, and cooling requirements compared to classical systems [Mansfield2025, Slysz2025], and the quantum technologies themselves vary in their environmental requirements. For example, quantum computers based on superconducting transmons and trapped ions demand vibration isolation, electromagnetic shielding, and temperature stability, whereas photonic systems have strict requirements for optical alignment, dust control, and humidity ranges. If co-located in HPC centers, these requirements must be satisfied without compromising the reliability of existing supercomputing operations. It is expected that redundant environmental infrastructure systems, such as power, chilled water, HVAC, vacuum, and cryosystems will help mitigate prolonged system downtimes. When HPC and SRQC machines are co-located, these environmental controls may be integrated into the existing facility management and monitoring systems used for DOE supercomputers, ensuring that both classical and quantum resources can be managed under a unified operational framework.

Operational readiness will require new procedures and workforce competencies. Quantum systems must undergo periodic calibration and maintenance that differ from typical HPC systems. This will require coordination among facility operators, system administrators, the user community, and hardware vendors. Establishing automated recalibration pipelines, telemetry feedback loops with responsive scheduling, and scheduled maintenance windows will help maximize uptime consistent with DOE service-level expectations. The identification of replaceable modular components in SRQC-scale machines is a research problem in its own right at the moment that the DOE should work with vendors to resolve.

The integration of quantum computers into DOE's computing infrastructure must include secure and transparent access mechanisms. Quantum systems should appear as managed resources within existing job schedulers and authentication frameworks, allowing users to submit and monitor hybrid workflows in terms of metadata, system context, and workflow tracking. Developing and standardizing these interfaces across facilities will reduce operational complexity and accelerate the transition from experimental hosting to production-class hybrid computing environments.

Finally, to realize this capability in its fullest, DOE should deploy one or more SRQCs (and follow-on PSQC machines), which can be installed in the facilities described above. This procurement will necessitate deep, ongoing public-private partnerships that culminate in a deployment. The process could begin with a request for information (RFI) in the first half of 2026 to determine vendor requirements and to gather the necessary information to begin any required facilities work. The RFI could be followed by a solicitation issued later in 2026. Pursuing multiple machines with distinct underlying technologies would reduce risks and bottlenecks associated with reliance on a single technology.

# 6. Transformative Acceleration of Scientific Quantum Utility

DOE can change the trajectory and landscape of quantum computing—accelerating the advent of unambiguous, broad-based quantum scientific utility from the 2030s to as soon as 2028—by supporting focused, goal-driven, high-priority R&D in three critical limiting areas: scientific applications, quantum algorithms, and logical architectures. All three are highly technical challenges that offer clear potential for transformative breakthroughs that could make scientific utility attainable on an SRQC. Goal-driven investment in applications, algorithms, and architectures represents a moon-shot opportunity for DOE to grasp scientific and technological leadership in quantum computing while accelerating scientific discovery across missions.

A multi-megaquop FTQC with an advanced architecture capable of performing $10^4$–$10^5$ hard logical operations on 100+ logical qubits will be able to perform high-fidelity quantum computations that vastly surpass simulatability by any known (or imaginable) classical computer; the earliest claims of such "quantum supremacy" date back to 2019 [Arute2019]. But it is not yet known whether a megaquop machine can be used to solve *useful* scientific problems. The gap between simulatability and scientific usefulness is not fully understood. Bridging this divide represents a primary pathway to achieving true quantum scientific utility with an SRQC as early as 2028–29. An opportunity exists for focused research efforts to close this gap by 2028. Success in this endeavor would be transformative, changing the landscape of quantum computing. In this section, we describe high-impact research directions that could enable breakthroughs and massively increase the impact of the SRQC deployment.

Three areas of research provide opportunities for transformative breakthroughs:

1. Discovery of new **scientific applications** amenable to FTQCs and aligned with DOE's mission.

2. Invention of new **quantum algorithms** that use logical resources more efficiently.

3. Creation of new **logical architectures** that use physical qubits more efficiently.

Each of these topics has a long history and is recognized as an essential pillar of quantum computing. However, each remains the specialized province of relatively few experts. Just as historical surges of focused effort in, for example, nuclear physics or rocketry supported by U.S. government science agencies produced rapid, transformative advances, these areas contain opportunities for breakthroughs that would emerge from focused research efforts. DOE is well positioned to assemble its leading experts and establish initiatives to advance these critical priorities.

## 6.1 Opportunities for Transformative Scientific Applications

Outperforming current or future HPC systems on known impactful scientific applications—for example, finding energies of FeMoco or the Hubbard model—is expected to require millions or billions of hard logical operations [Campbell2021, Babbush2023, Low2025]. In contrast, random circuit sampling (almost universally regarded as a contrived and unuseful problem) requires only thousands [Arute2019]. It is possible that this gap is not fundamental, but represents classical habits of thought in the definition of "impactful scientific applications." Historically, transformative scientific instruments such as the microscope, x-ray imaging, or radio telescopes have revealed new questions that were not previously imagined. It is reasonable to expect that once utility-scale quantum computers become available, they will catalyze the discovery or articulation of new scientific questions that are better adapted to quantum computing—and that, in retrospect, might have been solved with an SRQC.

There is, in principle, no reason why these questions could not be articulated before a DOE SRQC becomes available, just as microorganisms could, in principle, have been conjectured before the microscope. One suggestive, although insufficient, example is the dynamics of chemical or material systems. With classical methods, dynamics are harder to compute than energies, so scientists have focused on energies as representative properties. In contrast, on a quantum computer, dynamics may actually be easier than energy estimation, but fewer applications of dynamical simulation have been conceived and articulated because decades of pre-quantum-computing research focused on energies. We note the recent investigation of out-of-time-order correlation functions [Google2025b] as a promising example of the sort of speculative search for novel applications that might yield problems with radically lower resource costs.

## 6.2 Opportunities for Transformative Quantum Algorithms

Historically, the efficiency of classical algorithms (e.g., those for eigensystems) has improved steadily, producing efficiency gains in computation that have at times outpaced the improvement in hardware due to Moore's Law. Quantum algorithms are a far younger field but have thus far shown even more rapid improvement. It is taken for granted among experts that current quantum algorithms are far from optimal. Since resource costs have already been reduced by several orders of magnitude over the past 10 years, it is widely expected that additional orders of magnitude can be extracted from quantum algorithms for a range of problems—possibly even for Hamiltonian simulation (discussed in Section 3).

However, quantum algorithms for FTQCs—and the techniques and advancements that achieved order-of-magnitude speedups over the past 10–20 years—are extraordinarily technical and challenging. Relatively few scientists are experts in these techniques, and relatively little funding has supported their work. Extensive investment in the related field of NISQ and variational quantum algorithms laid the groundwork for the discovery of new FTQC algorithms and produced a trained workforce that can now be brought to bear on the harder problem of improving FTQC algorithms. Algorithm advancements are more likely to be solved by hard, well-informed teamwork among top experts in classical algorithms for DOE science problems and experts in the unique structural properties of quantum algorithms.

Opportunities for transformative advances include research that could (1) identify new algorithms or modifications to known algorithms that are particularly well suited to resource-constrained FTQCs, (2) discover or create novel connections between quantum algorithms and specific domain applications, (3) discover completely new quantum algorithms for known or innovative DOE science applications, (4) create and leverage novel and highly efficient quantum data types (e.g., logical fermions), (5) discover highly efficient replacements for ubiquitous but inefficient encodings (e.g., as block encodings of Hamiltonians improved on Trotterization), and (6) replace common subroutines with more efficient quantum-specific implementations (e.g., residue arithmetic).

## 6.3 Opportunities for Transformative Logical Architectures

For almost 15 years, most roadmaps to large-scale FTQCs have been based on a QEC code called the surface code [Dennis2002, Fowler2012a]. Although the surface code is conceptually straightforward and has a high error threshold, it is resource inefficient: several hundred physical qubits will be needed to encode a single logical qubit with sufficiently low error [Gidney2025]. Recent work in quantum low-density parity check (qLDPC) codes has led to the first explicit practical fault-tolerant architecture with a significantly smaller overhead (24:1) [Yoder2025] at the cost of more complex logical operations. In principle, much more efficient fault-tolerant protocols are possible, but the discovery and implementation of such protocols are complex, labor-intensive processes that require coordinated research by multidisciplinary teams of experts.

Logical architecture breakthroughs offer an opportunity to drastically accelerate the timetable for quantum utility, multiplying advances in applications and algorithms. Whereas better algorithms enable users to extract more powerful computations out of the same number of logical qubits and operations, better architectures enable a quantum computer to provide more logical qubits and logical operations with the same physical hardware. A 2028 quantum computer with 10,000 physical qubits and a physical error of $3 \times 10^{-4}$ per gate using the surface code would yield an SRQC offering 100 logical qubits and $10^7$ quops. With better QEC codes, routing, and implementations of hard gates, the same machine might provide 600 logical qubits and $10^{10}$ quops, transforming it into a PSQC that could solve the Hubbard model at a scientifically groundbreaking scale.

Examples of (hypothetical) transformative logical architecture breakthroughs include (1) discovery of a scalable family of qLDPC codes whose logical/physical encoding rate is greater than 1/12; (2) systematic creation of block codes that enable easy manipulation of block-encoded logical qubits; (3) significant reductions in the size of factories for magic-state distillation and cultivation; (4) invention of pairs of codes that enable simple, low-overhead code switching to implement universal logical gates; (5) creation of QEC codes whose structure enables reduced control and decoding bandwidth (e.g., because they are partially self-correcting); and (6) efficient logical operations co-designed specifically for certain qubit hardware (e.g., movable atomic lattices or photonic resource states).

# 7. Criteria for Success

By 2028, DOE should deploy SRQCs that demonstrate fault-tolerant quantum computing capable of performing science-relevant calculations and benchmarks; demonstrate progress toward PSQCs; have a workforce, staff, and facilities fully prepared to enable the use of SRQCs for scientific campaigns; and have made focused efforts to make utility achievable on an SRQC through transformative research.

## 7.1 What a Successful SRQC Would Look Like

At a minimum, SRQCs must be able to support compilation and execution of mini-applications relevant to users from each of the three main domain sciences described in Section 3. Programs that reach the maximum capacity of the SRQC's capability (both logical qubits and logical operations) represent the intended use of these systems and should be prioritized in design.

The SRQCs must perform scientific computations that, without necessarily exceeding the capabilities of a conventional computer, (1) demonstrate the viability of FTQCs for solving scientific problems and (2) clearly demonstrate a problem-size-dependent scaling that, when extrapolated to larger problems, predicts superclassical performance for next-generation PSQCs.

- Example 1: Computations of the ground-state energy density (to some precision, not necessarily impressive) for Fermi-Hubbard models on lattices of size $2 \times 2$, $4 \times 4$, and $6 \times 6$.

- Example 2: Performance of one, two, and four Trotter iterations for QPE on some encoding of a transition-metal molecule (e.g., FeS).

- Example 3: Computation of energy-energy correlations in pure SU(3) lattice gauge theory in two space and one time dimension on spatial lattices of size $3 \times 3$, $4 \times 4$, and $5 \times 5$.

The SRQCs will also be used to begin or accelerate the development of the software ecosystem necessary for FTQC systems, as described in Section 5. At a minimum, a compiler and associated tools will be required to make the machine minimally functional. Additionally, interfaces to the hardware devices to manage and interrogate the system will be needed to ensure smooth operations.

The software and applications that leverage SRQCs will evolve throughout the lifetime of the machine, potentially intersecting with several upgrades of the SRQC on the path to the final deployment of the PSQC machines. Where possible and reasonable, the system software stack will develop abstractions that are common to all the FTQC architectures deployed by DOE to enable a common, standard set of interfaces.

## 7.2 What Success in Transformative Research Would Look Like

The goal of the transformative research program outlined in Section 6 is to accelerate the advent of quantum scientific utility by making it achievable on SRQCs. Any of the following three accomplishments would constitute clear and compelling success of the program:

- A new **application** (scientific problem) of interest is discovered or invented that can be solved on an SRQC-class machine and is widely agreed to constitute useful domain science.

- A new quantum (or hybrid quantum-classical) **algorithm** is created that can solve a recognized problem of scientific interest that exceeds the capability of HPC and can be run on an SRQC-class machine.

- New logical **architecture** elements are invented that increase the quantum computational capability of an SRQC-class machine (defined now by physical hardware properties) to enable a quantum algorithm that achieves quantum scientific utility to be executed on the same hardware.

Enabling the SRQC to outperform exascale HPC on a recognized science problem by combining partial success of these goals would also constitute success.

Although the research described in Section 6 is high risk in nature, it would establish DOE's leadership role in quantum computing applications, algorithms, and architectures, paving the way for successful FTQC deployments in later years.

# 8. Next Steps and Roadmap

To enable demonstration of a scientific-scale FTQC in 2028, DOE must follow an aggressive schedule to organize the quantum computing and domain-science experts in the national laboratory complex, select vendors, and begin a robust co-design process. This process will include joint projects with the vendor community for R&D in hardware and runtime innovations, intermediate representations, and compiler interfaces. Co-design efforts aimed at deployment in the DOE ecosystem will also explore uptime metrics and facility targets for vendors and DOE centers.

The primary objectives of this project include the following:

- **Deploy** SRQCs that can execute at least $10^4$–$10^5$ hard logical operations on at least 100 logical qubits.

- **Develop** a portable, end-to-end software stack for execution, including compilation, circuit optimization, and orchestration of potential science workflows across the quantum-HPC ecosystem.

- **Advance** foundational research in FTQC technology for improved algorithms and applications, error-correcting codes, FTQC protocols, and logical architecture designs.

- **Establish** long-term quantum industry partnerships for co-design and software standardization.

- **Strengthen** partnerships with U.S. industry and academia to promote joint problem-solving, open standards, reproducibility, and workforce development.

- **Prepare** DOE facilities for operational hosting of quantum systems, including environmental, access, and security frameworks.
- **Demonstrate** quantum viability on key DOE applications (e.g., energy, materials, chemistry, and HENP).
- **Establish** benchmarks and success criteria for demonstrating quantum scientific utility by 2028.

The following milestones (Table 1) are proposed as phased activities to facilitate the deployment of SRQCs for DOE in 2028.

**Table 1. Milestones and activities for SRQC deployment.**

| Milestones and Activities |
|---|
| **Planning and Early R&D**<br>● Form a cross-lab steering committee<br>● Define architecture requirements and application drivers |
| **Prototype Integration and Application Co-Design**<br>● Issue RFIs to industry for quantum supercomputing concepts<br>● Examine existing designs and begin developing quantum-classical orchestration software stacks<br>● Examine existing designs and begin co-designing innovative logical architectures with vendors that can improve the implementation of known quantum algorithms for applications in materials, chemistry, and HENP<br>● Examine existing designs and begin co-designing innovative quantum algorithms for applications in materials, chemistry, and HENP deployed on known logical architectures<br>● Enhance workforce training and upskilling programs to prepare for SRQCs<br>● Initiate plans to deploy SRQCs at DOE laboratories<br>**Issue Solicitation**<br>● Issue a solicitation for SRQC systems based on the co-designed architecture, application, and software stack<br>● Select vendors; negotiate and finalize contracts<br>● Define an evaluation metrics framework |
| **Facility Readiness**<br>● Begin facilities preparations for SRQCs<br>**Establish Centers of Excellence for Vendor Progress Assessment/Collaboration**<br>● Begin coordinated testing of early-access SRQC prototypes and calibration pipelines<br>● Evaluate progress through data-driven modeling, benchmarking, and metrics assessment; conduct downselection reviews if necessary |
| **SRQC Demonstration**<br>● Begin deployment of large quantum hardware and control systems<br>● Deploy SRQCs with roughly 100 logical qubits<br>● Execute system verification and validation based on functional and performance criteria<br>● Conduct limited-access early-science experiments to validate system software, error correction, and hybrid workflows<br>● Release benchmark and performance results to inform next-generation design |
| **Toward PSQCs**<br>● Evaluate scientific outcomes from SRQC demonstrations<br>● Refine co-design models and facility standards based on operational experience<br>● Initiate planning for next-generation FTQC systems ($10^3$–$10^4$ logical qubits) targeting the early 2030s<br>● Expand training, documentation, and community engagement across DOE labs and user programs |

# 9. Glossary

**AI: Artificial intelligence**
A branch of computer science focused on systems capable of learning, reasoning, and pattern recognition. In quantum and quantum-HPC contexts, AI currently supports optimization, error correction, and exploration of large configuration spaces in hybrid workflows.

**API: Application programming interface**
A defined set of commands, data structures, and protocols enabling software components, such as quantum compilers, schedulers, or workflow managers, to interoperate across classical and quantum computing environments.

**Campaign (scientific campaign)**
A coordinated, multistep computational or experimental effort to address a defined scientific objective, often combining HPC simulations, quantum subroutines, data analysis, and AI-driven exploration under a unified workflow. Such workflows may operate in timescales that may go up to hours or days.

**Classical computing**
Conventional digital computation performed on information carriers that cannot exist in superposition states. It is usually implemented with binary logic and transistor-based architectures. Classical computers execute deterministic or probabilistic operations on bits, whereas quantum computers operate on qubits, which can support superposition and entanglement.

**DOE Office of Science computing facilities: ALCF, NERSC, OLCF**
DOE Office of Science user facilities that provide open, peer-reviewed access to some of the world's most powerful supercomputers for large-scale scientific research. The three DOE computing facilities—the Argonne Leadership Computing Facility (ALCF), the Oak Ridge Leadership Computing Facility (OLCF), and the National Energy Research Scientific Computing Center (NERSC)—enable computational campaigns that address grand scientific challenges across energy, materials, climate, and fundamental physics. In the context of this Blueprint, DOE's Office of Science computing facilities represent the operational and architectural model for integrating future quantum supercomputers into the national high-performance computing ecosystem.

**Fault tolerance**
A system's ability to tolerate faults in its operation by preventing any errors that occur from propagating catastrophically. This is typically achieved through redundancy. Everything from bridges to computers to governmental systems can be designed in a fault-tolerant manner. A system can be either small or large and still be fault tolerant.

**FTQC: Fault-tolerant quantum computer**
A quantum computer that performs computations on logical qubits that are protected from noise and errors by being encoded within a larger quantum substrate (usually many physical qubits) and periodically checked and corrected to eliminate errors.

**Hard logical operation**
A logical operation such as a *T* or Toffoli gate on 1–3 logical qubits that, because of how quantum error correction digitizes information and restricts the performance of arbitrary gates, can only be performed using a large number of elementary quops. Hard logical operations are the "long pole" of quantum algorithms, and their number is used to quantify the resource cost of a given algorithm instance. Innovations in logical architecture could change which operations are considered hard and which are considered easy. In this document, hard logical operations are usually assumed to be reliable.

**HENP: High-energy and nuclear physics**
The field of physics that investigates elementary particles and fields and their interactions.

**HPC: High-performance computing**
The use of massively parallel computing systems composed of thousands to millions of processing elements to perform large-scale numerical simulations, data analysis, and modeling tasks. In the DOE context, *HPC* refers to leadership-class supercomputing resources that provide the computational backbone for scientific discovery. Within this Blueprint, HPC also denotes the classical computing tier that complements quantum systems through orchestration, pre- and postprocessing, and hybrid quantum-classical workflows.

**ISA: Instruction set architecture**
The abstract machine-level interface that defines the set of logical operations and programming conventions supported by a quantum processor. This is analogous to classical ISAs but designed for quantum gates, qubit connectivity, and error-corrected operations.

**Logical qubit**
An effective qubit whose state is distributed across multiple subsystems. QEC codes exploit redundancies in logical qubit encodings to suppress subsystem errors through active feedback. In this document, logical qubits are usually assumed to be reliable.

**NISQ: Noisy intermediate-scale quantum**
The paradigm of performing quantum algorithms on logical qubits that are not encoded into QEC codes but are instead implemented at a 1:1 rate directly in physical qubit systems.

**PSQC: Production-scale quantum computer**
A large-scale, fault-tolerant quantum supercomputer (≥1,000 logical qubits, $\geq 10^9$ hard logical operations) capable of performing scientifically meaningful calculations beyond classical reach and targeted for DOE deployment in the early-to-mid 2030s.

**QEC: Quantum error correction**
A method for correcting errors in quantum systems. The most common form is syndrome-based QEC, in which one first measures an error syndrome on one or more logical qubits. The syndrome reveals information only about errors that may have occurred and nothing about the information stored in the logical qubits. A decoder then processes the syndrome and identifies a recovery operation intended to undo the error's effects. The recovery itself need not be acted upon immediately but merely tracked until it is necessary to deploy a corrective action, such as adapting future computational actions on logical qubits as needed.

**Quop: Quantum operation**
An elementary operation on logical qubit(s). What is considered elementary depends on the logical ISA. For example, a single-control, arbitrary-target controlled-NOT (CNOT) gate (also called a fanout-CNOT, or FCNOT) is considered elementary in lattice-surgery-based surface-code ISAs. The execution time of elementary gates scales only as a small multiple of the code distance of the logical qubit(s). Hard operations, such as logical $T$ or Toffoli gates, might require 100–1,000 quops when the time to distill special quantum states that realize them, called *magic states*, is considered. That said, hard logical operations can often be pipelined in logical ISAs so that many hard logical operations can occur in the time of one or a few quops via logical teleportation from logical magic states.

**Reliable**
A component or operation within a computational system is said to be reliable if it performs correctly with sufficiently high probability that the overall system or computation can perform its task successfully. The quantitative threshold for reliability is context dependent. For example, in a computation containing $10^6$ individual operations, "reliable" operations can fail with probability $\sim 10^{-6}$, but in a computation containing $10^9$ operations, "reliable" demands that each operation fail with a probability no greater than $10^{-9}$.

**SRQC: Scientifically relevant quantum computer**
A mid-scale fault-tolerant quantum system (~100 logical qubits) capable of executing tens to hundreds of thousands of reliable operations. SRQCs validate system integration, early software frameworks, and early scientific applications ahead of full utility-scale systems.

**Teraquop**
One trillion elementary quops, generally indicating a quantum computer capable of executing 1 trillion fault-tolerant logical idles (multiple cycles of fault-tolerant syndrome extraction) on individual logical qubits. For example, 1 billion cycles of fault-tolerant syndrome extraction on 1,000 logical qubits.

# 10. References

Aktar2025 S. Aktar et al., "Quantum Data Learning of Topological-to-Ferromagnetic Phase Transitions in the 2+1D Toric Code Loop Gas Model," arXiv (2025), https://doi.org/10.48550/arXiv.2511.16851.

Alam2025a F. Alam et al., "Fermionic Dynamics on a Trapped-Ion Quantum Computer Beyond Exact Classical Simulation," arXiv (2025), https://doi.org/10.48550/arXiv.2510.26300.

Alam2025b F. Alam et al., "Programmable Digital Quantum Simulation of 2D Fermi-Hubbard Dynamics Using 72 Superconducting Qubits," arXiv (2025), https://doi.org/10.48550/arXiv.2510.26845.

Alexeev2024 Y. Alexeev et al., "Quantum-Centric Supercomputing for Materials Science: A Perspective on Challenges and Future Directions," *Future Generation Computer Systems* 160 (2024): 666–710, https://doi-org.ornl.idm.oclc.org/10.1016/j.future.2024.04.060.

Alexeev2025 Y. Alexeev et al., "A Perspective on Quantum Computing Applications in Quantum Chemistry Using 25–100 Logical Qubits," *Journal of Chemical Theory and Computation* 21, no. 22 (2025): 11335–57, https://doi.org/10.1021/acs.jctc.5c01038.

Ameri2023 A., Ameri et al., "Quantum algorithm for the linear Vlasov equation with collisions", Proceedings of the 2023 IEEE International Conference on Quantum Computing and Engineering (QCE) (2023).

Aoki2024 Y. Aoki et al., *FLAG Review 2024*, arXiv (2024), https://doi.org/10.48550/arXiv.2411.04268.

Arute2019 F. Arute et al., "Quantum Supremacy Using a Programmable Superconducting Processor," *Nature* 574, no. 7779 (2019): 505–10, https://doi.org/10.1038/s41586-019-1666-5.

Asai2023 S. Asai et al., *Exploring the Quantum Universe: Pathways to Innovation and Discovery in Particle Physics*, report of the Particle Physics Project Prioritization Panel (2023), https://www.usparticlephysics.org/2023-p5-report/.

Assi2024 B. Assi and H. Lamm, "Digitization and Subduction of $SU(N)$ Gauge Theories," *Physical Review D* 110 (2024): 074511, https://doi.org/10.1103/PhysRevD.110.074511.

Babbush2023 R. Babbush et al., "Quantum Simulation of Exact Electron Dynamics Can Be More Efficient Than Classical Mean-Field Methods," *Nature Communications* 14 (2023): 4058, https://doi.org/10.1038/s41467-023-39024-0.

Barata2022 J. Barata et al., "Medium Induced Jet Broadening in a Quantum Computer," *Physical Review D* 106, no. 7 (2022): 074013, https://doi.org/10.1103/PhysRevD.106.074013.

Bauer2016 B. Bauer et al., "Hybrid Quantum-Classical Approach to Correlated Materials," *Physical Review X* 6, no. 3 (2016): 031045, https://doi.org/10.1103/PhysRevX.6.031045.

Bauer2020 B. Bauer et al., "Quantum Algorithms for Quantum Chemistry and Quantum Materials Science," *Chemical Reviews* 120, no. 22 (2020): 12685–717, https://doi.org/10.1021/acs.chemrev.9b00829.

Bauer2023 C. Bauer et al., "Quantum Simulation for High-Energy Physics," *PRX Quantum* 4, no. 2 (2023): 027001, https://doi.org/10.1103/PRXQuantum.4.027001.

BaySmidt2025 A. J. Bay-Smidt et al., "Fault-Tolerant Simulation of Generalized Hubbard Models," arXiv (2025), https://doi.org/10.48550/arXiv.2501.10314.

Beck2019 D. Beck et al., *Nuclear Physics and Quantum Information Science*, report from the NSAC Quantum Information Science Subcommittee, U.S. Department of Energy Office of Science (2019), https://science.osti.gov/-/media/np/pdf/Reports/NSAC_QIS_Report.pdf.

Beck2023 D. Beck et al., “Quantum Information Science and Technology for Nuclear Physics. Input into US Long-Range Planning, 2023,” arXiv (2023), https://doi.org/10.48550/arXiv.2303.00113.

Beck2024 T. Beck et al., “Integrating Quantum Computing Resources into Scientific HPC Ecosystems,” *Future Generation Computer Systems* 161 (2024): 11–25, https://doi.org/10.1016/j.future.2024.06.058.

Bellonzi2024 N. Bellonzi et al., “Feasibility of Accelerating Homogeneous Catalyst Discovery with Fault-Tolerant Quantum Computers,” arXiv (2024), https://doi.org/10.48550/arXiv.2406.06335.

Bluvstein2025 D. Bluvstein et al., “Architectural Mechanisms of a Universal Fault-Tolerant Quantum Computer,” arXiv (2025), https://doi.org/10.48550/arXiv.2506.20661.

Bouland2019 A. Bouland et al., “On the Complexity and Verification of Quantum Random Circuit Sampling,” *Nature Physics* 15, no. 2 (2019): 159–63, https://doi.org/10.1038/s41567-018-0318-2.

Bravyi2022 S. Bravyi et al., “The Future of Quantum Computing with Superconducting Qubits,” arXiv (2022), https://doi.org/10.48550/arXiv.2209.06841.

Buchs2025 G. Buchs et al., “The Role of Quantum Computing in Advancing Scientific High-Performance Computing: A Perspective from the ADAC Institute,” arXiv (2025), https://doi.org/10.48550/arXiv.2508.11765.

Burgholzer2025 L. Burgholzer et al., “The Munich Quantum Software Stack: Connecting End Users, Integrating Diverse Quantum Technologies, Accelerating HPC,” arXiv (2025), https://doi.org/10.48550/arXiv.2509.02674.

Byrnes2006 T. Byrnes and Y. Yamamoto, “Simulating Lattice Gauge Theories on a Quantum Computer,” *Physical Review A* 73 (2006): 022328, https://doi.org/10.1103/PhysRevA.73.022328.

Caldwell2025 S. A. Caldwell et al., “Platform Architecture for Tight Coupling of High-Performance Computing with Quantum Processors,” arXiv (2025), https://doi.org/10.48550/arXiv.2510.25213.

Campbell2021 E. Campbell, “Early Fault-Tolerant Simulations of the Hubbard Model,” *Quantum Science and Technology* 7, no. 1 (2021): 015007, https://doi.org/10.1088/2058-9565/ac3110.

Camps2024 D. Camps et al., “Evaluation of the Classical Hardware Requirements for Large-Scale Quantum Computations,” *ISC High Performance 2024 Research Paper Proceedings* (2024), https://doi.org/10.23919/isc.2024.10528937.

Camps2025 D. Camps et al., “Quantum Computing Technology Roadmaps and Capability Assessment for Scientific Computing -- An Analysis of Use Cases from the NERSC Workload,” arXiv (2025), https://doi.org/10.48550/arXiv.2509.09882.

Carlson2018 J. Carlson et al., *Quantum Computing for Theoretical Nuclear Physics, A White Paper Prepared for the U.S. Department of Energy, Office of Science,*

*Office of Nuclear Physics*, U.S. Department of Energy Office of Science (2018), https://doi.org/10.2172/1631143.

Ciavarella2021 A. N. Ciavarella, N. Klco, and M. J. Savage, "Trailhead for Quantum Simulation of SU(3) Yang-Mills Lattice Gauge Theory in the Local Multiplet Basis," *Physical Review D* 103, no. 9 (2021): 094501, https://doi.org/10.1103/PhysRevD.103.094501.

Ciavarella2024 A. N. Ciavarella and C. W. Bauer, "Quantum Simulation of SU(3) Lattice Yang-Mills Theory at Leading Order in Large-$N_c$ Expansion," *Physical Review Letters* 133, no. 11 (2024): 111901, https://doi.org/10.1103/PhysRevLett.133.111901.

Ciavarella2025 A. N. Ciavarella, I. M. Burbano, and C. W. Bauer, "Efficient Truncations of SU($N_c$) Lattice Gauge Theory for Quantum Simulation," *Physical Review D* 112, no. 5 (2025): 054514, https://doi.org/10.1103/ylqb-phv5.

Cohen2021 T. D. Cohen, H. Lamm, S. Lawrence, and Y. Yamauchi, "Quantum Algorithms for Transport Coefficients in Gauge Theories," *Physical Review D* 104, no. 9 (2021): 094514, https://doi.org/10.1103/PhysRevD.104.094514.

Dalzell2023 A. M. Dalzell et al., "Quantum Algorithms: A Survey of Applications and End-to-End Complexities," arXiv (2023), https://doi.org/10.48550/arXiv.2310.03011.

Dasu2025 S. Dasu et al., "Breaking Even with Magic: Demonstration of a High-Fidelity Logical Non-Clifford Gate," arXiv (2025), https://doi.org/10.48550/arXiv.2506.14688.

Davoudi2021 Z. Davoudi et al., "Nuclear Matrix Elements from Lattice QCD for Electroweak and Beyond-Standard-Model Processes," *Physics Reports* 900 (2021): 1–74.

Davoudi2022 Z. Davoudi et al., "Report of the Snowmass 2021 Topical Group on Lattice Gauge Theory," arXiv (2022), https://doi.org/10.48550/arXiv.2209.10758.

Dennis2002 E. Dennis et al., "Topological Quantum Memory," *Journal of Mathematical Physics* 43, no. 9 (2002): 4452–505, https://doi.org/10.1063/1.1499754.

DeTar2004 C. DeTar and S. Gottlieb, "Lattice Quantum Chromodynamics Comes of Age," *Physics Today* 57, no. 45 (2004), https://doi.org/10.1063/1.1688069.

DiMeglio2024 A. Di Meglio et al., "Quantum Computing for High-Energy Physics: State of the Art and Challenges," *PRX Quantum* 5, no. 3 (2024): 037001, https://doi.org/10.1103/PRXQuantum.5.037001.

DOE2024 *2024 Quantum Information Science Applications Roadmap* (2024), https://science.osti.gov/-/media/QIS/pdf/DOE_QIS_Roadmap_Final.pdf.

Egan2021 L. Egan et al., "Fault-Tolerant Control of an Error-Corrected Qubit," *Nature* 598, no. 7880 (2021): 281–86, https://doi.org/10.1038/s41586-021-03928-y.

Farrar2017 G. R. Farrar, "Stable Sexaquark," arXiv (2017), https://doi.org/10.48550/arXiv.1708.08951.

Farrar2020 G. R. Farrar, Z. Wang, and X. Xu, "Dark Matter Particle in QCD," arXiv (2020), https://arxiv.org/pdf/2007.10378.

Feynman1982 R. Feynman, "Simulating Physics with Computers," *International Journal of Theoretical Physics* 21, nos. 6/7 (1982), https://s2.smu.edu/~mitch/class/5395/papers/feynman-quantum-1981.pdf.

Fowler2012a A. G. Fowler et al., "Surface Codes: Towards Practical Large-Scale Quantum Computation," *Physical Review A* 86, no. 3 (2012): 032324, https://doi.org/10.1103/PhysRevA.86.032324.

Fowler2012b A. G. Fowler, "Time-Optimal Quantum Computation," arXiv (2012), https://doi.org/10.48550/arXiv.1210.4626.

Gheorghiu2025 V. Gheorghiu and M. Mosca, "Quantum Resource Estimation for Large Scale Quantum Algorithms," *Future Generation Computer Systems* 162 (2025): 107480, https://doi.org/10.1016/j.future.2024.107480.

Gidney2021 C. Gidney et al., "A Fault-Tolerant Honeycomb Memory," *Quantum* 5, no. 605 (2021), https://doi.org/10.22331/q-2021-12-20-605.

Gidney2025 C. Gidney, "How to Factor 2048 Bit RSA Integers with Less Than a Million Noisy Qubits," arXiv (2025), https://doi.org/10.48550/arXiv.2505.15917.

Goings2022 J. J. Goings et al., "Reliably Assessing the Electronic Structure of Cytochrome P450 on Today's Classical Computers and Tomorrow's Quantum Computers," *Proceedings of the National Academy of Sciences* 119, no. 38 (2022), https://doi.org/10.1073/pnas.2203533119.

Google2025a Google Quantum AI, "Quantum Error Correction Below the Surface Code Threshold," *Nature* 638, no. 8052 (2025): 920–26, https://doi.org/10.1038/s41586-024-08449-y.

Google2025b Google Quantum AI, "Observation of Constructive Interference at the Edge of Quantum Ergodicity," *Nature* 646, no. 8086 (2025): 825–30, https://doi.org/10.1038/s41586-025-09526-6.

Gottesman2016 D. Gottesman, "Quantum Fault Tolerance in Small Experiments," arXiv (2016), https://doi.org/10.48550/arXiv.1610.03507.

Harrigan2024 M. P. Harrigan et al., "Expressing and Analyzing Quantum Algorithms with Qualtran," arXiv (2024), https://doi.org/10.48550/arXiv.2409.04643.

Hogan2025 N. Hogan et al., "A Quantum Computing Approach to Efficiently Simulating Correlated Materials Using Impurity Models and Dynamical Mean Field Theory," arXiv (2025), https://arxiv.org/html/2508.05738v1.

Illa2024 M. Illa, C. E. P. Robin, and M. J. Savage, "Qu8its for Quantum Simulations of Lattice Quantum Chromodynamics," *Physical Review D* 110, no. 1 (2024): 014507, https://doi.org/10.1103/PhysRevD.110.014507.

Joó2019 B. Joó et al., "Status and Future Perspectives for Lattice Gauge Theory Calculations to the Exascale and Beyond," *The European Physical Journal A* 55, no. 11 (2019): 199, https://doi.org/10.1140/epja/i2019-12919-7.

Joseph2023 I. Joseph et al., "Quantum computing for fusion energy science applications", *Physics of Plasmas* 30, 010501 (2023)

Kan2021 A. Kan and Y. Nam, “Lattice Quantum Chromodynamics and Electrodynamics on a Universal Quantum Computer,” arXiv (2021), https://doi.org/10.48550/arXiv.2107.12769.

Keen2020 T. Keen et al., “Quantum-Classical Simulation of Two-Site Dynamical Mean-Field Theory on Noisy Quantum Hardware,” *Quantum Science and Technology* 5 (2020): 35001, https://iopscience.iop.org/article/10.1088/2058-9565/ab7d4c/ampdf.

Kharazi2026 T. Kharazi et al., “A Sublinear-Time Quantum Algorithm for High-Dimensional Reaction Rates”, arXiv:2601.15523 (2026).

Kim2023 Y. Kim et al., “Evidence for the Utility of Quantum Computing Before Fault Tolerance,” *Nature* 618, no. 7965 (2023): 500–505, https://doi.org/10.1038/s41586-023-06096-3.

Kogut1975 J. B. Kogut and L. Susskind, “Hamiltonian Formulation of Wilson’s Lattice Gauge Theories,” *Physical Review D* 11 (1975): 395–408, https://doi.org/10.1103/PhysRevD.11.395.

Kürkçüoglu2025 D. M. Kürkçüoglu, H. Lamm, O. Ogunkoya, and L. Pierattelli, “The 2T-Quoctit: A Two-Mode Bosonic Qudit for High Energy Physics,” arXiv (2025), https://doi.org/10.48550/arXiv.2509.21941.

Lee2025 K. Lee, F. Turro, and X. Yao, “Quantum Computing for Energy Correlators,” *Physical Review D* 111, no. 5 (2025): 054514, https://doi.org/10.1103/PhysRevD.111.054514.

Lin2024 J. Lin, D. Luo, X. Yao, and P. E. Shanahan, “Real-Time Dynamics of the Schwinger Model as an Open Quantum System with Neural Density Operators,” *Journal of High Energy Physics* 2024 (2024): 211, https://doi.org/10.1007/JHEP06(2024)211.

Liu2021 J.P. Liu et al., “Efficient quantum algorithm for dissipative nonlinear differential equations”, *Proc. Nat. Acad. Sci.* 118 (35) e2026805118 (2021)

Lloyd1996 S. Lloyd, “Universal Quantum Simulators,” *Science* 273, no. 5278 (1996): 1073–78, https://doi.org/10.1126/science.273.5278.1073.

Low2019 G. H. Low and I. L. Chuang, “Hamiltonian Simulation by Qubitization,” *Quantum* 3 (2019): 163, https://doi.org/10.22331/q-2019-07-12-163.

Low2025 G. H. Low et al., “Fast Quantum Simulation of Electronic Structure by Spectral Amplification,” *Physical Review X* 15, no. 4 (2025): 041016, https://doi.org/10.1103/pb2g-j9cw.

Mansfield2025 E. Mansfield et al., “First Practical Experiences Integrating Quantum Computers with HPC Resources: A Case Study With a 20-Qubit Superconducting Quantum Computer,” arXiv (2025), https://doi.org/10.48550/arXiv.2509.12949.

McArdle2020 S. McArdle et al., “Quantum Computational Chemistry,” *Reviews of Modern Physics* 92, no. 1 (2020): 015003, https://doi.org/10.1103/RevModPhys.92.015003.

Nie2024 X. Nie et al., “Self-Consistent Determination of Single-Impurity Anderson Model Using Hybrid Quantum-Classical Approach on a Spin Quantum Simulator,” *Physical Review Letters* 133 (2024): 140602, https://doi.org/10.1103/PhysRevLett.133.140602.

Paetznick2024 A. Paetznick et al., “Demonstration of Logical Qubits and Repeated Error Correction with Better-Than-Physical Error Rates,” arXiv (2024), https://doi.org/10.48550/arXiv.2404.02280.

Perez2025 S. O. Perez, E. M. Murairi, E. J. Gustafson, and H. Lamm, “Primitive Quantum Gates for an SU(3) Discrete Subgroup: Σ(72×3),” arXiv (2025), https://doi.org/10.48550/arXiv.2511.17437.

Postler2022 L. Postler et al., “Demonstration of Fault-Tolerant Universal Quantum Gate Operations,” *Nature* 605, no. 7911 (2022): 675–80, https://doi.org/10.1038/s41586-022-04721-1.

Proctor2025 T. Proctor et al., “Benchmarking Quantum Computers,” *Nature Reviews Physics* 7, no. 2 (2025): 105–18, https://doi.org/10.1038/s42254-024-00796-z.

Reichardt2024a B. W. Reichardt et al., “Demonstration of Quantum Computation and Error Correction with a Tesseract Code,” arXiv (2024), https://doi.org/10.48550/arXiv.2409.04628.

Reichardt2024b B. W. Reichardt et al., “Fault-Tolerant Quantum Computation with a Neutral Atom Processor,” arXiv (2024), https://doi.org/10.48550/arXiv.2411.11822.

Reiher2017 M. Reiher et al., “Elucidating Reaction Mechanisms on Quantum Computers,” *Proceedings of the National Academy of Sciences* 114, no. 29 (2017): 7555–60, https://doi.org/10.1073/pnas.1619152114.

Rhodes2024 M. Rhodes et al., “Exponential Improvements in the Simulation of Lattice Gauge Theories Using Near-Optimal Techniques,” *PRX Quantum* 5 (2024): 040347, https://doi.org/10.1103/PRXQuantum.5.040347.

Rhodes2026 M. L. Rhodes, S. Pathak, and R. W. Chien, “Quantum Simulation of Lattice Gauge Theories Coupled to Fermionic Matter via Anyonic Regularization,” arXiv (2026), https://doi.org/10.48550/arXiv.2603.15820.

Rubin2024 N. Rubin et al., "Quantum computation of stopping power for inertial fusion target design", *Proc. Nat. Acad. Sci.* 121 (23) e2317772121 (2024)

Slysz2025 M. Slysz et al., “Hybrid Classical-Quantum Supercomputing: A Demonstration of a Multi-User, Multi-QPU and Multi-GPU Environment,” arXiv (2025), https://doi.org/10.48550/arXiv.2508.16297.

Shehata2025 A. Shehata et al., “Bridging Paradigms: Designing for HPC-Quantum Convergence,” arXiv (2025), https://doi.org/10.1016/j.future.2025.107980.

Shehata2026 A. Shehata et al., “Quantum-HPC Software Stacks and the openQSE Reference Architecture: A Survey,” arXiv (2026), https://doi.org/10.48550/arXiv.2604.20912.

Siew2026 R. X. Siew, S. Chandrasekharan, and T. Bhattacharya, “Continuum Limit of a Qubit-Regularized SU(3) Lattice Gauge Theory with Glueballs,” arXiv (2026), https://doi.org/10.48550/arXiv.2603.01215.

Steckmann2023 T. Steckmann et al., “Mapping the Metal-Insulator Phase Diagram by Algebraically Fast-Forwarding Dynamics on a Cloud Quantum Computer,” *Physical Review Research* 5, no. 2 (2023): 023198, https://doi.org/10.1103/PhysRevResearch.5.023198.

Than2025 A. T. Than et al., “Observation of Quantum-Field-Theory Dynamics on a Spin-Phonon Quantum Computer,” arXiv (2025), https://doi.org/10.48550/arXiv.2509.11477.

Turro2024 F. Turro, A. Ciavarella, and X. Yao, “Classical and Quantum Computing of Shear Viscosity for (2+1)D SU(2) Gauge Theory,” Physical Review D 109, no. 11 (2024): 114511, https://doi.org/10.1103/PhysRevD.109.114511.

Turro2025 F. Turro and X. Yao, “Emergent Hydrodynamic Mode on SU(2) Plaquette Chains and Quantum Simulation,” *Physical Review D* 111, no. 9 (2025): 094502, https://doi.org/10.1103/PhysRevD.111.094502.

Vorwerk2022 C. Vorwerk et al., “Quantum Embedding Theories to Simulate Condensed Systems on Quantum Computers,” *Nature Computational Science* 2, no. 7 (2022): 424–32, https://doi.org/10.1038/s43588-022-00279-0.

Yoder2025 T. J. Yoder et al., “Tour de Gross: A Modular Quantum Computer Based on Bivariate Bicycle Codes,” arXiv (2025), https://doi.org/10.48550/arXiv.2506.03094.

Yoshioka2024 N. Yoshioka et al., “Hunting for Quantum-Classical Crossover in Condensed Matter Problems,” *npj Quantum Information* 10 (2024): 45, https://doi.org/10.1038/s41534-024-00839-4.

# Appendix A. Quantum Computing R&D Roundtable Findings

March 4–5, 2026, Rockville, MD

## A.1 Context

The National Laboratory members of the DOE Laboratory Quantum Supercomputing Blueprint Team invited leading researchers to a roundtable meeting in March 2026 to identify critical technical gaps, recommend near-term R&D priorities, and align the community on actions needed to maximize the scientific value of the systems described in the Blueprint Report. This appendix is the synthesis of these discussions by the National Laboratory Organizing Committee (NLOC); it is not a direct reporting of attendee contributions.

Roundtable participants were charged with answering a central question:

*How can the community deliver a clear win for domain-science problems of DOE interest on a scientifically relevant quantum computer (SRQC), namely one possessing approximately 100 logical qubits and able to execute approximately 100 million quops, by 2028?*

As noted in the main Blueprint Report, the NLOC acknowledged to roundtable participants that while SRQCs are well beyond the reach of what classical computers can simulate, the path to a scientific win with such machines is unclear and would benefit from input from the roundtable as to what investment choices, informed by judgment about prioritization, would be the best.

Participants considered several related questions:

- What priority R&D directions in logical architectures, algorithms, and their co-design will best facilitate this win?
- What priority R&D directions in materials science, quantum chemistry, and High Energy and Nuclear Physics (HENP) applications will best facilitate this win?
- What priority capability investments will best ensure that the outcomes of these R&D directions are deployed for a win?

After an introduction and a reading of the charge, DOE Under Secretary for Science Dr. Darío Gil made opening remarks and responded to a moderated Q&A session. A selected group of nine attendees then presented five-minute lightning talks conveying their initial thoughts. Over two full days, attendees then participated in breakouts and developed specific recommendations in five primary areas: algorithms, architectures, co-design, chemistry and materials applications, and HENP applications. They also came together collectively and identified cross-cutting themes and enabling infrastructure that would best support these recommendations. The remainder of this appendix summarizes the findings of this meeting.

## A.2 Cross-Cutting Themes

All five primary area breakout teams agreed that achieving a scientifically meaningful result on SRQCs will depend on close coordination across the full quantum computational stack. At the scale of ~100 logical qubits, they noted that performance will remain dominated by architectural constraints and execution overheads. Many argued that isolated advances will have limited significance within the 2028 timeframe. They suggested that, instead, tightly integrated workflows linking algorithm design, fault-tolerant execution, and hardware capabilities will define the primary path to progress.

Key themes include the following:

- **Co-design across algorithms and architectures has great potential.** Algorithms, compilation, resource states, and architectural constraints should be developed and evaluated together. This approach provides a strong possibility for resource reductions in certain scenarios.

- **Common representations are needed to enable comparison.** The lack of shared abstractions for architectures and workflows limits coordination and convergence.

- **Benchmarks and reference problems are important to drive progress.** Some domains have representative problems, whereas others lack well-defined benchmarks with clear observables and targets.

- **Progress depends on problem-driven, interdisciplinary workflows.** Advancing toward a 2028 demonstration will require highly committed and coordinated efforts across domain scientists, algorithm developers, software developers, and system experts. This should be balanced with efforts to build a robust, sustainable quantum computing ecosystem.

- **Investigating new approaches to fault tolerance could be a valuable area of focus to produce sought-after FTQC performance and resource gains.** Alternative approaches to error correction and detection, such as dynamical FTQC, may improve efficiency under early FTQC constraints.

## A.3 Primary Areas

### A-3.1 Quantum Logical Architectures

Participants identified four key challenges faced by quantum logical architecture R&D:

1. R&D for these architectures is highly fragmented and lacks common frameworks, languages, and software tools with which to innovate and explore. Researchers often work in isolation between institutions and comparing approaches can be difficult, obstructing hybridization of ideas and cross-platform adoption.

2. Verifying correctness, emulating behaviors, and benchmarking performances of logical architectures are hindered by a lack of sufficiently flexible and modular shared software tools. Vendors are not motivated to develop cross-platform solutions.

3. Slow hardware design cycles and bespoke near-term architectures limit the speed with which one can iteratively explore and deploy new logical architectures, such as those co-designed on a faster cycle with algorithms for DOE science applications.

4. The workforce size may be inadequate to innovate the breakthroughs required to achieve a scientific win by 2028 SRQCs.

This breakout group identified the following research priorities to address these challenges:

- **Cultivate a shared logical architecture R&D environment by fostering the creation of common frameworks, languages, and software tools.** Some potential building blocks discussed included Gottesman's dynamical FTQC framework [arXiv:2403.04163], the ZX calculus [arXiv:2012.13966], and various logical quantum intermediate representation (IR) languages. Shared libraries of published architectures expressed with these paradigms could foster hybridization of ideas with more rapid innovation cycles and enable reproducible comparison, interoperability, and reuse.

- **Coordinate the development of shared methods and tools for verifying, validating, and benchmarking FTQC architectures.** Some ideas suggested included developing a quantum "digital twin" framework, hosting a shared library of validated reference architectures, augmenting something like Stim [arXiv:2103.02202] to be a more comprehensive analysis tool, and creating better cross-platform resource estimation tools.

- **Ensure that DOE, in partnership with vendors, can deploy newly innovated logical architectures as firmware upgrades that obviate full hardware upgrades.** Noted examples of firmware upgradability include QEC codes, logical gate protocols, decoder algorithms, and logical data types, especially for any of these co-designed with DOE science applications in mind.

- **Invest in the growth of the workforce.** Ideas generated included working with universities on faculty hiring support and curriculum development, funding cross-training initiatives to draw in experts in related areas, and establishing scholarships and fellowships for students working in this area.

## A-3.2 Quantum Algorithms

Over the past 20+ years, steady incremental progress in quantum algorithms has reduced the complexity of scientifically relevant quantum computations by orders of magnitude. Nevertheless, there was agreement among participants that existing algorithmic approaches are likely insufficient to enable scientifically relevant calculations on quantum computers in 2028–29. Participants noted a dearth of algorithms tailored for early FTQCs, the need for better methods and tools for prototyping and evaluating algorithms, and the value of having agreed-upon reference applications and algorithms to spur and assess progress.

- **Algorithms suited for SRQCs.** Research in new methods for implementing quantum simulation, novel approaches to computing quantities of interest in quantum physical systems, and problems outside quantum simulation for which SRQCs may provide an advantage.

- **A living suite of benchmark problems.** The community would benefit from a collaborative public repository containing a range of well-defined target computations, covering multiple scientific domains, for SRQCs. The content for each computational target would ideally include (1) a concrete problem specification, including details of the Hamiltonians, initial states, observables to be measured, evolution times, and desired precision, for example; (2) an assessment of the best known classical approaches; (3) state-of-the-art quantum implementations and/or resource estimates for representative architectures.

- **Experiments to evaluate critical algorithmic techniques.** The arrival of SRQCs will enable quantum computational experiments that can inform and accelerate the development of algorithmic techniques in ways that theory and classical simulation cannot. The design of such computational experiments is nontrivial and should begin before the arrival of the hardware.

## A-3.3 Co-Design and Compilation

Quantum algorithms and logical architectures are often treated as distinct. Algorithms and applications are conceived abstractly and then compiled, routed, and implemented for specific universal architectures, whereas QEC codes and FTQC protocols are designed and optimized independently. However, exceptions to this rule, such as Gidney's designs for factoring [arXiv:2505.15917] or the STAR architecture [arXiv:2303.13181, arXiv:2509.18294], show that *co-design* of architectures and algorithms can reduce physical resource requirements. Many roundtable participants suggested that co-designing algorithms, QEC, and logical operations, especially for SRQC hardware, could accelerate the timeline for scientific utility. The promise of co-design will likely depend on physical architecture, but almost all participants agreed that co-optimizing gate-synthesis techniques on SRQCs with algorithms—often incorporated into compilation—could accelerate progress.

The roundtable participants discussed investigating whether, where, and how effectively algorithms and architectures can be co-optimized. Demonstrations that co-design works and proofs limiting its utility will each help direct future research. Concrete items include the following:

- **Quantum computational primitives of SRQCs:** Investigate methods for extra-efficient logical operations in specific near-term architectures. Investigate catalytic resource states that, once prepared, can reduce the cost of computation by, for example, enabling efficient single-qubit rotations [arXiv:2305.07720] or replacing costly operations like division with easier operations like multiplication [arXiv:2306.08585].

- **Resource estimation and compilation tools**: Create and disseminate better tools for resource estimation, especially compilation of complex logical operations into elementary ones.

- **Partial fault tolerance**: Investigate and systematize techniques for performing efficient logical operations in *partially fault-tolerant* ways, e.g., fault-tolerant compilation that improves the accuracy of specific observables or error mitigation designed for SRQC use.

- **Inherently fault-tolerant operations**: Investigate and develop complex (beyond Clifford + T) computational primitives with intrinsic error-detecting and error-correcting properties (as in [arXiv:1606.01904], for example) by synthesizing a CCZ gate using eight T gates while performing error detection. Methods agnostic to the underlying code would be the most powerful.

## A-3.4 Chemistry and Materials

Quantum chemistry and materials science applications have been at the forefront of demonstrating the benefits of quantum computing. The roundtable discussed challenge problems of interest, but emphasized that the field is broad with many problems of interest. Effort is needed by the community to accelerate workflows to enable rapid feedback with domain-science experts to investigate applications and algorithms. Domain problems benefit from horizontal integration between SMEs and quantum computing experts who can deploy quantum algorithms, QEC, and hardware-aware optimizations to significantly accelerate problem components on quantum computing systems. For example, bringing QEC approaches into algorithm design can result in a vast reduction of resources needed for the problem (e.g., [arXiv:2511.09124]). Similarly, bringing physical and chemical awareness and intuition into algorithm design has already been shown to reduce the resources required, and we can expect this to continue. This integration will also enable a proper resource estimation for chemistry and materials problems that account for the constraints.

The roundtable discussed application constraints and candidate applications in chemistry and material science.

**Chemistry**

Essential information about entanglement in quantum systems can be extracted directly during quantum simulations. Strong correlation effects in both ground and excited states relevant to catalysis, photocatalysis, and various spectroscopic processes will benefit from more accurate calculations on a 100 logical-qubit platform. One specific candidate application is the design of next-generation, energy-efficient metal-sulfide-based catalysis for fuel production and emission mitigation. Concrete developments include the following:

- Develop an improved description of strong correlation effects in the ground and excited states relevant to catalysis, photocatalysis, and spectroscopies in various energy regimes.

- Develop downfolding and embedding methods that encode missing correlation effects (dynamical correlation) that cannot be directly captured today in 100-qubit simulations.

- Devise improved downfolding and embedding methods by leveraging machine learning techniques.

**Material Science**
Fermionic lattice models are a promising domain area in which acceleration with quantum computing can bring improvements in accuracy and time-to-solution. Specifically, FTQCs can bring early scientific wins in the calculation of spectroscopic signatures of fermionic lattice models of quantum materials. These calculations are necessary to use experimental data to establish an underlying theory by determining the microscopic interactions and mechanisms that lead to materials behavior. An important takeaway is that it is important to identify problems that emphasize dynamics over ground states.

Microscopic interactions encoded in model Hamiltonians can be inferred from experimental spectroscopic data (e.g., inelastic neutron scattering, angle-resolved photoemission spectroscopy, and scanning tunneling microscopy). However, this requires accurate calculations of spectroscopic signatures in model Hamiltonians. Quantum embedding offers a scalable, physically meaningful pathway toward calculating these spectroscopic signatures. It reduces the complexity to something manageable on a 100-logical-qubit quantum computer by mapping the lattice problem onto an impurity problem embedded in a bath designed to represent the rest of the system. Impurity models have the advantage of being weighted toward the dynamical problem—the ground state is easier to prepare than it is for a straightforward lattice model, and this can be tuned as the size of the impurity is increased. Such embedding approaches enable a meaningful scientific problem to be run on a 100-qubit system, and readily extrapolate to larger systems, if and when they become available. Application domain targets of DOE relevance include Moiré superlattices and superconductivity in nickelates and cuprates. Concrete developments include the following:

- Develop embedding algorithms that use the FTQC infrastructure selectively and judiciously. What error rates are **really needed?** (For example, is $10^{-8}$ needed everywhere? Can we use low-precision operations for certain operations?)

- Develop methods to obtain the simplest Hamiltonians that qualitatively match experiments, for example, relative pole heights and locations (50% accuracy).

- Explore strategies to improve the quality of AI models by fine-tuning them with training data obtained on a quantum computer. Use this training data to improve models beyond the capabilities of classical computers. An advantage of this approach is that it can be developed and tested at present through coarse training with approximate methods (e.g., density functional theory) and fine-tuning with more accurate methods (e.g., tensor networks or coupled cluster).

## A-3.5 High Energy Physics and Nuclear Physics

Simulating the dynamics of matter under extreme conditions and at scale is required to optimize the scientific impact of the experimental, theoretical, and computational programs in HENP in the United States. Many such simulations lie beyond the capabilities of classical computation and quantum simulations are currently the only foreseeable technology capable of illuminating these phenomena. In addition to observation of complex dynamics, quantum simulations of these systems open the possibility of identifying new quantum signatures that will enhance the discovery potential of current and future experiments.

The roundtable breakout team noted that over the past decade, quantum simulations of (1+1)D systems have advanced significantly, constrained primarily by the limitations of current NISQ devices. However, dynamics in one spatial dimension differ qualitatively from those in higher dimensions. The advent of SRQCs will enable the exploration of gauge dynamics in (2+1)D—and ultimately (3+1)D—on modest volumes, establishing a foundation for scalable quantum simulations. Importantly, this would not be about building a "femtoscope" for a single problem but developing a capability to advance the HENP discovery program in multiple directions. Real-time simulations of strongly interacting, strongly correlated matter at subatomic scales will eventually impact our understanding of hadronization and hydrodynamics at the Large Hadron Collider (LHC), neutrino-nucleus interactions at the Deep Underground Neutrino Experiment (DUNE), and nuclear reactions relevant to energy and national security. These targets provide concrete opportunities to establish quantum advantage with first-generation SRQCs using simplified and approximate approaches.

**Research Directions:**

We anticipate measuring real-time observables relevant to hadronization and jet fragmentation, correlation functions relevant to transport coefficients, with finite-density systems as a longer-term target. Recent estimates for the FTQC resources needed to simulate the time evolution of SU(3) quantum fields reveal a broad landscape that is strongly dependent on the truncation strategy and system size. As a fiducial benchmark example, consider 50 Trotter steps of time evolution in an SU(3) pure gauge theory. In this case, results at modest system sizes that are nonetheless classically infeasible to simulate show that digitization choices drive resource requirements, making optimal representation a central challenge for near-term quantum simulation of non-Abelian gauge theories. At the same time, significant headroom remains for algorithmic improvements, including specialized distillation and catalysis techniques. Full-stack resource estimation remains underdeveloped, underscoring the need for closer collaboration with experts in resource modeling and fault-tolerant architecture.

Wide variations in resource estimates make clear that encoding and QEC strategies are central to practical quantum simulations. By tailoring QEC to the symmetries of the target theory, gauge redundancy becomes a feature rather than overhead, reducing costs and improving performance. Extending these approaches to non-Abelian theories offers a promising path toward scalable, hardware-aligned simulations of the Standard Model.

The success and acceleration of such efforts will depend on expanding and sustaining a skilled workforce spanning HENP, algorithms, software, systems integration, and QEC. The field faces a clear imbalance: the number of high-impact problems far exceeds the number of researchers prepared to tackle them, and continued progress will only widen this gap. Addressing it requires broad recruitment, deliberate cross-disciplinary training, and critically sustainable career paths, particularly in computational and quantum information fields. Creating stable, long-term opportunities is essential to attract, retain, and empower the talent needed to realize the program's scientific goals.

**Specific Research Priorities**:

- Dynamics of confinement: Develop and benchmark quantum algorithms for real-time evolution, emphasizing observables that probe the dynamics of confinement.

- (2+1)D models: Develop a systematic SRQC roadmap for confining non-Abelian gauge theories in going from (2+1)D to (3+1)D.
- Cross-domain program in HENP: Define a first set of SRQC-computable observables that are relevant to hadronization (LHC), neutrino-nucleus interactions (DUNE), and nuclear reaction dynamics.
- Quantum advantage pathway: Develop a resource optimization plan including encodings, truncation schemes, and circuit constructions in order to achieve FTQC advantage.
- HEP-aware QC: Investigate error correction and fault tolerance strategies specific to HENP examples and quantify reductions in resource requirements relative to generic approaches.
- Workforce development: Establish a sustained program at the HENP-QIS interface to train early career researchers and create long-term positions targeted to enhancing and maintaining domain-specific expertise.

# A.4 Enabling Infrastructure

In addition to the five primary topic areas, Roundtable participants highlighted infrastructural needs in software, data, hardware, and workforce coordination. These were identified as key enablers for the priority research directions outlined above and vital to achieving meaningful results on SRQCs.

1. **Integrated, Modular Software Stack:** A composable software stack is required to connect algorithms, compilation, QEC, and execution. Toolchains must be modular, allowing components such as encodings, compilers, and decoders to be developed, evaluated, and exchanged without rebuilding full systems. Programming environments should support rapid prototyping, implementation, and benchmarking with low user burden. The stack must support end-to-end workflows linking application definition, compilation, execution, and analysis, while integrating simulation and resource estimation capabilities.
2. **Shared Repositories and Standardized Artifacts:** Shared repositories are required to enable reproducible evaluation and comparison. These include device models, benchmark problems, reference implementations, validated FTQC architectures, and protocol libraries. Curated application libraries, benchmark circuits, and syndrome datasets support consistent evaluation across architectures and algorithms, whereas domain-specific libraries provide reusable components for common computational tasks.
3. **Resource Estimation, Simulation, and Validation Tools:** Tools are required to evaluate resource requirements and performance under realistic conditions. Resource estimation must be linked to compilation and execution to assess end-to-end costs. Classical simulation frameworks must support noise modeling, large-scale studies, and validation against experimental data. QEC decoders and fast simulators are needed to model system behavior.

4. **Scalable Compute and Hardware Access:** Access to HPC systems is required for simulation, optimization, and design-space exploration. Access to SRQC systems and emulators is required for testing and validation. These resources must support iterative workflows across simulation and execution within a hybrid computing environment.

5. **Coordinated, Cross-Disciplinary R&D:** Progress requires coordinated efforts across applications, algorithms, QEC, compilation, and hardware. Cross-disciplinary collaboration is necessary to address system-level challenges, including structured collaboration across institutions, sustained workforce development, and continued support for theory and simulation.

## Attendee List

| Subject Matter Experts | | | |
|---|---|---|---|
| Alexey Gorshkov | NIST | Karol Kowalski | PNNL |
| Andrew Baczewski | Sandia | Kristan Temme | IBM |
| Andrew Childs | UMD | Krysta Svore | NVIDIA |
| Andrew Cross | IBM | Alexander (Lex) Kemper | NC State |
| Andrew Potter | Quantinuum | Lin Lin | LBNL / UCB |
| Antonio Corcoles | IBM | Martin Savage | Washington |
| Bela Bauer | Microsoft | Natalie Klco | Duke |
| Christian Bauer | LBNL | Nathan Wiebe | PNNL |
| Daniel Gottesman | UMD | Nicholas (Nick) Rubin | Google |
| Daniel Litinski | PsiQuantum | Nicolas Delfosse | IonQ |
| Dave Bacon | Google | Sam Pallister | PsiQuantum |
| Henry (Hank) Lamm | Fermilab | Sergey Bravyi | IBM |
| Isaac (Ike) Chuang | MIT | Stephen Jordan | Google |
| Isaac Kim | UC Davis | Tanmoy Bhattacharya | LANL |
| Joonho Lee | Harvard | Thomas Maier | ORNL |
| | | Zohreh Davoudi | UMD |
| **Implementation** | | | |
| **NQISRC Reps / Leadership** | | **Organizing Committee/Moderators** | |
| Bert deJong (QSA) | LBNL | Andrew Landahl | Sandia |
| Travis Humble (QSC) | ORNL | Robin Blume-Kohout | Sandia |
| Anna Grassellino (SMQS) | FNL | Ryan Bennink | ORNL |
| Jennifer (Jen) Dionne (Q-NEXT) | Stanford | Arjun Shankar | ORNL |
| Nathan Wiebe (C2QA) | PNNL | Laura Schulz | ANL |
| | | Salman Habib | ANL |
| | | Hai Ah Nam | LBNL |
| | | Daan Camps | LBNL |

| Observers | | | |
|---|---|---|---|
| Pavel Lougovski | DOE | Hal Finkel | DOE |
| Tanner Crowder | DOE | Ben Brown | DOE |
| Marco Fornari | DOE | Enrique (Riq) Parra | DOE |
| Adam Kinney | DOE | Jordan Thomas | DOE |
| Sharon Stephenson | DOE | Resham Kulkarni | DOE |
| Tom Setterston | DOE | Rachael Al-Saadon | DARPA |